\documentclass[useAMS]{mn2e}
\usepackage{graphicx}
\usepackage{url}
\usepackage{multirow}
\usepackage{multicol}
\usepackage{caption}
\usepackage{float}
\usepackage{placeins}
\usepackage{tabularx}
\usepackage{amsmath}

\title[No AGN contribution to hydrogen reionization]{No evidence for a significant AGN contribution to cosmic hydrogen reionization}
\author[Parsa et al.]{Shaghayegh Parsa$^{1}$\thanks{E-mail:
shp@roe.ac.uk}, James S. Dunlop$^{1}$\thanks{E-mail:
jsd@roe.ac.uk}, Ross J. McLure$^{1}$\\
$^{1}$Institute for Astronomy, University of Edinburgh, Royal Observatory, Edinburgh, EH9 3HJ}

\begin{document}

\pagerange{\pageref{firstpage}--\pageref{lastpage}} \pubyear{2014}

\maketitle

\label{firstpage}

\begin{abstract}
We reinvestigate a claimed sample of 22 X-ray detected active galactic nuclei (AGN) at redshifts $z > 4$, which has reignited the debate as to whether young galaxies or AGN reionized the
Universe. These sources lie within the GOODS-S/CANDELS field, and we examine both the robustness of the claimed X-ray detections (within the {\it Chandra} 4Ms imaging) and perform an independent
analysis of the photometric redshifts of the optical/infrared counterparts. We confirm the reality of only 15 of the 22 reported X-ray detections, and moreover find that only 
12 of the 22 optical/infrared counterpart galaxies actually lie robustly at $z > 4$. Combining these results we find convincing evidence for only 7 X-ray AGN at $z > 4$ in the GOODS-S field, of which only one lies at $z > 5$. We recalculate the evolving far-UV (1500\,\AA) luminosity density produced by AGN at high redshift, and find that it declines rapidly 
from $z \simeq 4$ to $z \simeq 6$, in agreement with several other recent studies of the evolving AGN luminosity function. The associated rapid decline in inferred hydrogen-ionizing 
emissivity contributed by AGN falls an order-of-magnitude short of the level required to maintain hydrogen ionization at $z \simeq 6$. We conclude that all available evidence continues to favour a scenario in which young galaxies reionized the Universe, with AGN making, at most, a very minor contribution to cosmic hydrogen reionization.

\end{abstract}

\begin{keywords}
galaxies: evolution - galaxies: high-redshift - quasars: supermassive black holes - cosmology: theory - dark ages, reionization, first stars
\end{keywords}

\section {Introduction}

Our understanding of cosmic reionization has advanced steadily in recent years.
The latest microwave background measurements now indicate a (model dependent)
`mean' redshift for hydrogen reionization of $\langle z \rangle \simeq 7.8 - 8.8$ (Planck Collaboration 2016),
while observations of the Gunn-Peterson effect (Gunn \& Peterson 1965) in high-redshift quasar
spectra (e.g. Fan et al. 2006; Bolton et al. 2011; McGreer et al. 2015; Barnett et al. 2017),
and the decline in observable Lyman-$\alpha$ emission from high-redshift galaxies (e.g. Stark et al. 2010, 2015; 
Pentericci et al. 2011, 2014; Curtis-Lake et al. 2012;
Schenker et al. 2012, 2014; Treu et al. 2013; Tilvi et al. 2014; Dijkstra et al. 2014)
suggest the process was essentially complete by $z \simeq 6$.

Through parallel efforts, our knowledge of early galaxy evolution has
been pushed back to within $\simeq 0.5$\,Gyr of the Big Bang
(see Dunlop 2013, Madau \& Dickinson 2014, and Stark 2016 for reviews charting this progress),
with the latest high-redshift galaxy surveys now enabling measurement of the
evolving rest-frame UV galaxy luminosity function out to at least $z \simeq 10$
(e.g. Ellis et al. 2013; Dunlop et al. 2013; McLure et al. 2013; Oesch et al 2014; Bowler et al.
2014, 2015, 2017a; Bouwens et al. 2015a; McLeod et al. 2015, 2016; Parsa et al. 2016; Ishigaki et al. 2017).

As a consequence, it now appears highly likely that the rapidly growing population of early
star-forming galaxies could indeed have bathed the Universe in sufficient
high-energy photons to produce/maintain
cosmic hydrogen reionization on a timescale consistent with the
aforementioned observational constraints (Robertson et al. 2010, 2013, 2015; Bouwens et al. 2015b).
However, two key uncertainties remain to be resolved. First, analyses combining all available
constraints indicate that most of the relevant `action' has yet to be discovered, with
the ionizing photon budget dominated by low-luminosity galaxies undetected
by the {\it Hubble Space Telescope} ({\it HST}) (although see Sharma et al. 2016).
Second, the ionizing photon escape fraction from the high-redshift galaxies
is generally required to be $\simeq 10 - 20$\% to achieve timely reionization (e.g. Robertson et al. 2013, 2015;  Fontanot et al. 2014; Finkelstein et al. 2015;
although see Weisz \& Boylan-Kolchin 2017)
significantly higher than typically found for lower-redshift star-forming analogues (e.g.
Siana et al. 2010; Grazian et al. 2017).

Arguably these are not serious problems, and further observational advances
can be expected from 2019 with the
{\it James Webb Space Telescope} ({\it JWST\/}), given its potential to
detect rest-frame UV emission out to $z \simeq 30$. Moreover, there are good theoretical
reasons to expect typical ionizing-photon escape fractions to grow at high redshift (e.g.
Sharma et al. 2017), and problems related to escape fraction
(or, equivalently, the duration of hard ionizing radiation from young star-forming galaxies)
may be further aleviated by the inclusion of binary stars in spectral synthesis
models of galaxies (Eldridge, Izzard \& Tout 2008; Eldridge \& Stanway 2009; Stanway et al. 2016; Bowler et al. 2017b).

Nonetheless, such concerns, coupled with the knowledge that supermassive
black-holes are certainly known to exist at $z > 7$ (Mortlock et al. 2011),
continue to generate interest in the possibility that
active galactic nuclei (AGN) could have played a significant role in cosmic hydrogen reionization.

Given that the accretion process within AGN generates a hard radiation spectrum, with
a potentially large escape fraction, AGN are obviously excellent sources of ionizing photons,
and the key issue is whether they are sufficiently numerous at early times to contribute
significantly to the hydrogen reionziation process. Indeed, there is very little debate that
quasars/AGN are responsible for He reionization, which observations indicate is completed by
$z \simeq 2.5 - 3$, and requires a radiation field with $h\nu > 54.4$\,eV photons
(e.g. Miralda-Escude et al. 2000; McQuinn et al. 2009; Haardt \& Madau 2012; Compostella et al. 2013).
This follows empirically from the fact that the emissivity of quasars at $z \simeq 3$ is likely sufficient to reionize HeII atoms (Furlanetto \& Oh 2008; Faucher-Giguere et al. 2008;
Haardt \& Madau 2012). 

However, the number density of bright quasars and AGNs declines at $z>3$ (Fan et al. 2000; Richards et al. 2006; Jiang et al. 2009, 2016;  Willott et al. 2010; Glikman et al. 2011; Fontanot et al. 2012;
Masters et al. 2012; McGreer et al. 2013; Akiyama et al. 2017) and, using parametric 
extrapolations for the form/evolution of the AGN luminosity function (LF),  Haardt \& Madau (2012) estimated that the ionizing emissivity of AGNs at $z \simeq 6$ was only 
$\simeq 10^{23}\,{\rm erg\,s^{-1}Hz^{-1} Mpc^{-3}}$, an order-of-magnitude smaller than required to maintain hydrogen reionization in the intergalactic medium (IGM) at that redshift.  

Nevertheless, it can be argued that this issue is still not closed because of remaining uncertainty over the faint-end slope of the 
AGN LF at high redshifts. Attempts to better measure the faint end slope of the optical quasar LF at $z \simeq 3-4$ indicate that it remains relatively 
flat, with $\beta \simeq 1.5$, similar to measurements at lower redshift (Siana et al. 2008; Glikman et al. 2011). Attempts have also been made to extend measurement of the X-ray LF of AGN
to high redshifts, utilising {\it XMM} and {\it Chandra} imaging of the COSMOS field (Brusa et al. 2010; Civano et al. 2011) and, more recently, the 4\,Ms {\it Chandra} imaging in the  Chandra Deep Field South (CDF-S) survey (Xue et al. 2011). In particular, Fiore et al. (2012) adopted a new approach to try to exploit the full power of the 4\,Ms {\it Chandra} CDFS imaging to uncover high-redshift
AGN, by searching for X-ray detections at the positions of known high-redshift galaxies in the deep optical--near-IR imaging within the central GOODS-S field. However, at the time
of that study, the full CANDELS imaging (Grogin et al. 2011) of GOODS-S had yet to be completed, and although Fiore et al. (2012) reported the discovery of several 
new X-ray AGN at $z > 3$, a meaningful determination of the faint-end slope of the AGN X-ray LF at higher redshifts still proved elusive.

Motivated by this approach, following completion of the CANDELS WFC3/IR imaging 
(Guo et al. 2013), Giallongo et al. (2015) undertook a new search for high-redshift X-ray sources in GOODS-S,  and reported
the detection of a substantial population of faint X-ray AGN at $z > 4$, which, they claimed, could have produced sufficient co-moving far-UV emissivity, even at $z \simeq 6$, to maintain hydrogen
ionization in the IGM. Specifically, Giallongo et al. (2015) reported the discovery of 22 X-ray AGN at $z > 4$ in the GOODS-S field (only 8 of which featured in the purely X-ray selected {\it Chandra} 4\,Ms catalogue of Xue et al. 2011), and derived a new enhanced estimate of the comoving ionizing emissivity of AGN at early times, based on a new determination of the 
far-UV ($\lambda_{rest} = 1450$\,\AA) LF of AGN over the redshift range $4 < z < 6$.
This work has inspired theorists to reappraise the potential role of AGN in cosmic reionization (e.g. Madau \& Haardt 2015; Qin et al. 2017; Mitra, Choudhury \& Ferrara 2018).

 \begin{table*}
\caption{The positions and spectroscopic+photometric redshift information for the 22 $z > 4$ X-ray AGN candidates selected by Giallongo et al. (2015). $ID_{G15}$, $ID_{X11}$ and $ID_{3D-HST}$ represent respectively the identification number for each source as published in the Giallongo et al. (2015), Xue et al. (2011) and Skelton et al. (2014) catalogues. 3D-HST, G15 and H15 indicate the photometric redshifts of the optical/near-IR sources as determined by Skelton et al. (2014), Giallongo et al. (2015) and Hsu et al. (2014). The photometric redshifts determined in this study 
are then given under the heading `This Work', followed in the final three columns by the $\chi^2$ value achieved in the SED fit, the SED type T (G for Galaxy, A for AGN) of the best fitting
template, and finally the dust reddening of the best-fitting template, $E(B-V)$, assuming the dust attenuation law of Calzetti et al. (2000).}
\resizebox{0.85\textwidth}{!}{\begin{minipage}{\textwidth}
\begin{tabular}{c|c|c|c|c|c|c|c|c|c|c|c|c}
\hline
  \multicolumn{3}{c|}{ID}  &
  \multirow{2}{*}{RA} &
  \multirow{2}{*}{Dec} &
  \multirow{2}{*}{$z_{spec}$} &
  \multicolumn{4}{c|}{$z_{phot}$}  &
  \multirow{2}{*}{$\chi^{2}$} &
  \multirow{2}{*}{T} &
  \multirow{2}{*}{$E(B-V)$\,/\,mag} \\
  \cline{1-3}\cline{7-10}
  G15 & X11 & 3D-HST & & & & 3D-HST & G15 & H14 & This Work & & \\
\hline
  273 & 403 & 471 & 53.1220463 & $-$27.9387409 & 4.76 & 4.76 & 4.49  & 4.70  & 4.25 & 12.49 & G & 0.5\\
  4285 & - & 7895 & 53.1664941 & $-$27.8716803 &- & 4.32 & 4.28  & - & 0.46 & 6.31 & G & 0.0\\
  4356 & 485 & 8052 & 53.1465968 & $-$27.8709872 &- & 5.12 & 4.70 & 2.47 &  3.62 & 0.42 & A & 0.6\\
  4952 & - & - & 53.1605007 & $-$27.8649890 &- & 4.32 & 4.32  & - &  4.34 & 14.77 & A & 0.0\\
  5375 & 331 & 10474 & 53.1026292 & $-$27.8606307 & - & 4.37 & 4.41  & 0.70  & 0.23 & 1.79 & A & 0.5\\
  5501 & - & 10764 & 53.1280240& $-$27.8593930 & - & 5.76 & 5.39  & - & 3.96 & 4.14 & G & 0.5\\
  8687 & - & 17724 & 53.0868634& $-$27.8295859&- & 4.43 & 4.23 &  - & 3.57 & 7.97 & G & 0.1\\
  8884 & - & 18119 & 53.1970699 & $-$27.8278566 & - & 4.76 & 4.52 & - & 4.40 & 3.95 & G & 0.5\\
  9713 & - & 19509 & 53.1715890 & $-$27.8208052 & 5.70 & 5.76 & 5.86 &  - & 5.97 & 10.64 & G & 0.1\\
  9945 & - & 19906 & 53.1619508 & $-$27.8190897& 4.48 & 4.43 & 4.34  & - & 4.28 & 8.44 & G & 0.1\\
  11287 & - & 22175 & 53.0689924 & $-$27.8071692 & - & 4.82 & 4.94 &  - & 4.90 & 5.02 & G & 0.2\\
  12130 & - & 23582 & 53.1514304&$-$27.7997601& 4.62 & 4.54 & 4.43 &  - & 4.39 & 18.17 & G & 0.1\\
  14800 & - & 27960 & 53.0211735 & $-$27.7823645 & 4.82 & 4.88 & 4.92 &  - & 4.80 & 1.23 & G & 0.1\\
  16822 & 371 & 31374 & 53.1115637 & $-$27.7677714 & - & 4.82 & 4.52 & 3.24 &  4.50 & 2.32 & G & 0.4\\
  19713 & 392 & 36223 & 53.1198898 & $-$27.7430349 &- & 5.49 & 4.84  & 2.4 & 2.88 & 1.29 & G & 0.8\\
  20765 & 521 & 37989 & 53.1583449 & $-$27.7334854 & - & 3.15 & 5.23  & 2.59 &  1.17 & 2.44 & A & 0.7\\
  23757 & - & 41824 & 53.2036444& $-$27.7143907 & - & 3.86 & 4.13  & - & 0.14 & 4.52 & A & 0.4\\
  28476 & - & - & 53.0646867 & $-$27.8625539 & - & - & 6.26 &  - &  6.16 & 1.41 & G & 0.7\\
  29323 & 156 & 15927 & 53.0409764 & $-$27.8376619 &- & 5.24 & 9.73 & 4.65 & 1.21 & 1.27 & A & 1.0\\
  31334 & - & 27976 & 53.2131871 & $-$27.7816486 &- & 0.80 & 4.73 & - & 4.86 & 4.78 & A & 0.0\\
  33073 & - & 37292 & 53.0547529 & $-$27.7368325 & - & 4.76 & 4.98 & - & 4.93 & 0.05 & G & 0.0\\
  33160 & 85 & 37803 & 53.0062504 & $-$27.7340678 & - & 5.06 & 6.06  & 3.36  & 2.52 & 1.02 & A & 0.7\\
\hline\end{tabular}
\end{minipage}}
\end{table*}

However, these results have proved somewhat controversial.
First, while the detection of additional X-ray sources with the aid of near-infrared positional priors is not unexpected, the application of a similar technique by Cappelluti et al. (2016)
to the same datasets found significantly fewer high-redshift X-ray detections than claimed by Giallongo et al. (2015). In addition, again utilising essentially the same {\it Chandra} 4\,Ms, {\it HST} CANDELS and {\it Spitzer} imaging, Weigel et al. (2015) failed to find convincing evidence for any X-ray source at $z > 5$ within GOODS-S. 
Moreover, recent studies of the X-ray AGN LF (Aird et al. 2015; Fotopoulou et al. 2016)  out to $z \simeq 4-5$ suggest that the faint-end slope of the 
total X-ray AGN LF stays unchanged, or if anything flattens with increasing redshift. Most recently, Vito et al. (2016) used the new 7\,Ms {\it Chandra} imaging in the CDF-S, coupled with the CANDELS optical+near-IR imaging, to measure the total X-ray emission from 2076 galaxies in GOODS-S over the redshift range $3.5 < z < 6.5$. 
They used stacking methods and derived the first meaningful constraints on the faint-end slope of the X-ray LF at $z > 4$, again finding evidence for fairly flat slopes, consistent 
with previous studies at comparable redshifts (Vito et al. 2014; Georgakakis et al. 2015). Futhermore,
Vito et al. (2016) found no evidence of significant X-ray emission in galaxy stacks at redshifts higher than  $z \simeq 5$. 

In an attempt to contribute to this debate, and help clarify the origin of the some of the disagreements outlined above, we have undertaken a reanalysis of the 22 $z > 4$ X-ray AGN 
reported by Giallongo et al. (2015). We have determined new photometric redshifts for the proposed 
optical+near-IR counterparts, and have reassessed the robustness of the proposed X-ray detections. 
Based on the revised sample of high-redshift AGN we have then re-estimated the high-redshift evolution the far-UV comoving emissivity produced by AGN, and the consequent  
potential contribution of AGN to cosmic hydrogen reionization.
  
The layout of the rest of this paper is as follows. In Section 2 we summarize the available X-ray and supporting optical+near-IR data, and the sample of 22 objects discussed by 
Giallongo et al. (2015). Then, in Section 3 we describe our new determination of the photometric redshifts of the optical+near-IR objects, and present a reassessment of the reliability 
of the claimed X-ray detections in the 4\,Ms {\it Chandra} imaging. In Section 4 we use the resulting revised sample of high-redshift AGN in tandem with results from brighter AGN surveys
to re-determine the likely form and evolution of the high-redshift far-UV LF of AGN, and hence the contribution of AGN to hydrogen ionizing emissivity at early times.
We discuss our results in the context of other recent studies in Section 5, and summarize our conclusions in Section 6. Throughout, all magnitudes are quoted in the AB
system (Oke 1974; Oke \& Gunn 1983), and all cosmological calculations assume a flat cosmology with 
$\Omega_m = 0.3$, $\Omega_\Lambda = 0.7$ and H$_0$ = 70 km s$^{-1}$ Mpc$^{-1}$.

\section{Data}
\subsection{X-ray Data}

The X-ray data utilised here is that provided by the {\it Chandra} X-ray Observatory through the first 4\,Ms of imaging undertaken within 
the Chandra Deep Field South (CDF-S) survey (Xue et al. 2011). The publicly-available X-ray imaging, spanning the energy range $0.5 - 10$\,keV,  was reduced following the procedures 
described in Luo et al. (2008).

\subsection{Optical and near-IR data}

In this study we have used the publicly-available {\it HST} WFC3/IR and {\it HST} ACS imaging of the Great Observatories Origins Deep Survey South (GOODS-S) field provided by the Cosmic Assembly Near-Infrared Deep Extragalactic Legacy Survey (CANDELS) (Grogin et al. 2011; Koekemoer et al. 2011; Windhorst et al. 2011), and the associated pre-existing {\it HST} optical (Giavalisco et al. 2004), ground-based VLT $U$-band (Nonino et al. 2009) and $K_s$-band (Retzlaff et al. 2010; Fontana et al. 2014), and {\it Spitzer} IRAC imaging (Ashby et al. 2013), as summarized by Guo
et al. (2013). The sources were detected, and their iso-photal fluxes in the $HST$ bands measured using \textsc{sextractor} v2.8.6 (Bertin \& Arnouts 1996) 
in dual-image mode, with $H_{160}$ as the detection image. As described in Guo et al. (2013), in this field the Template FITting (\textsc{tfit}) method (Laidler et al. 2007) has been applied to generate the matched photometry from the lower angular resolution $U$, $K_{s}$ and IRAC imaging. The GOODS-S catalogue provided by Guo et al. (2013) contains 34930 sources within a sky 
area of 173\,arcmin$^{2}$, with photometry in the $U$, $B_{435}$, $V_{606}$, $i_{775}$, $z_{850}$, $Y_{098}$, $Y_{105}$, $J_{125}$, $H_{160}$, $K_{s}$, $IRAC_{3.6{\mu}m}$ and $IRAC_{4.5{\mu}m}$ bands. 

\subsection{The Giallongo et al. (2015) sample}

To extract their new sample of proposed high-redshift $z > 4$ X-ray AGN, Giallongo et al. (2015) first isolated a sample of $z > 4$ optical/near-IR sources from within the 
Guo et al. (2013) GOODS-S catalogue, using the photometric redshifts assembled by the CANDELS team (Dahlen et al. 2013). This resulted in a sample of 1113 sources in the CANDELS/GOODS-S region with 
photometric or spectroscopic redshifts $z > 4$. They then used the positions of these proposed $z > 4$ sources as derived from the {\it HST} WFC3/IR $H_{160}$ imaging as 
priors in the procedure proposed by Fiore et al. (2012) to search for significant X-ray flux at these input positions. In essence, Giallongo et al. (2015) pushed the sensitivity limit of the
{\it Chandra} imaging by looking for clustering of X-ray counts in space–time-energy parameter space. The result was a claimed sample of 22 significant X-ray detections from
within the parent sample of 1113 $z > 4$ sources. This sample is presented in Table\,1. As noted by Giallongo et al. (2015), 8 of these 22 sources were already reported in the
X-ray selected catalogues of Xue et al. (2011) and Hsu et al. (2014), with a further 3 previously reported by Fiore et al. (2012), albeit with somewhat different photometric redshifts.
Thus, 14 of these 22 sources were newly `discovered' with the aid of the $H_{160}$ input prior positions, and are presumably therefore too faint to have been discovered utilising straightforward established X-ray source detection techniques applied purely to the {\it Chandra} imaging.

Table 1 summarizes the available information on these sources (prior to the present study) including catalogue numbers from Giallongo et al. (2015), Xue et al. (2011), and 3D-HST 
(Skelton et al. 2014), 
$H_{160}$ positions, spectroscopic redshifts (available for 5 optical/near-IR sources) and photometric redshifts as derived by Giallongo et al. (2015), Hsu et al. (2014), and the 
3D-HST team (Skelton et al. 2014). This table also contains our own new photometric redshift determinations which we discuss and present in the next Section.

\begin{figure*}
\centering
\includegraphics[width=\textwidth, height=600pt]{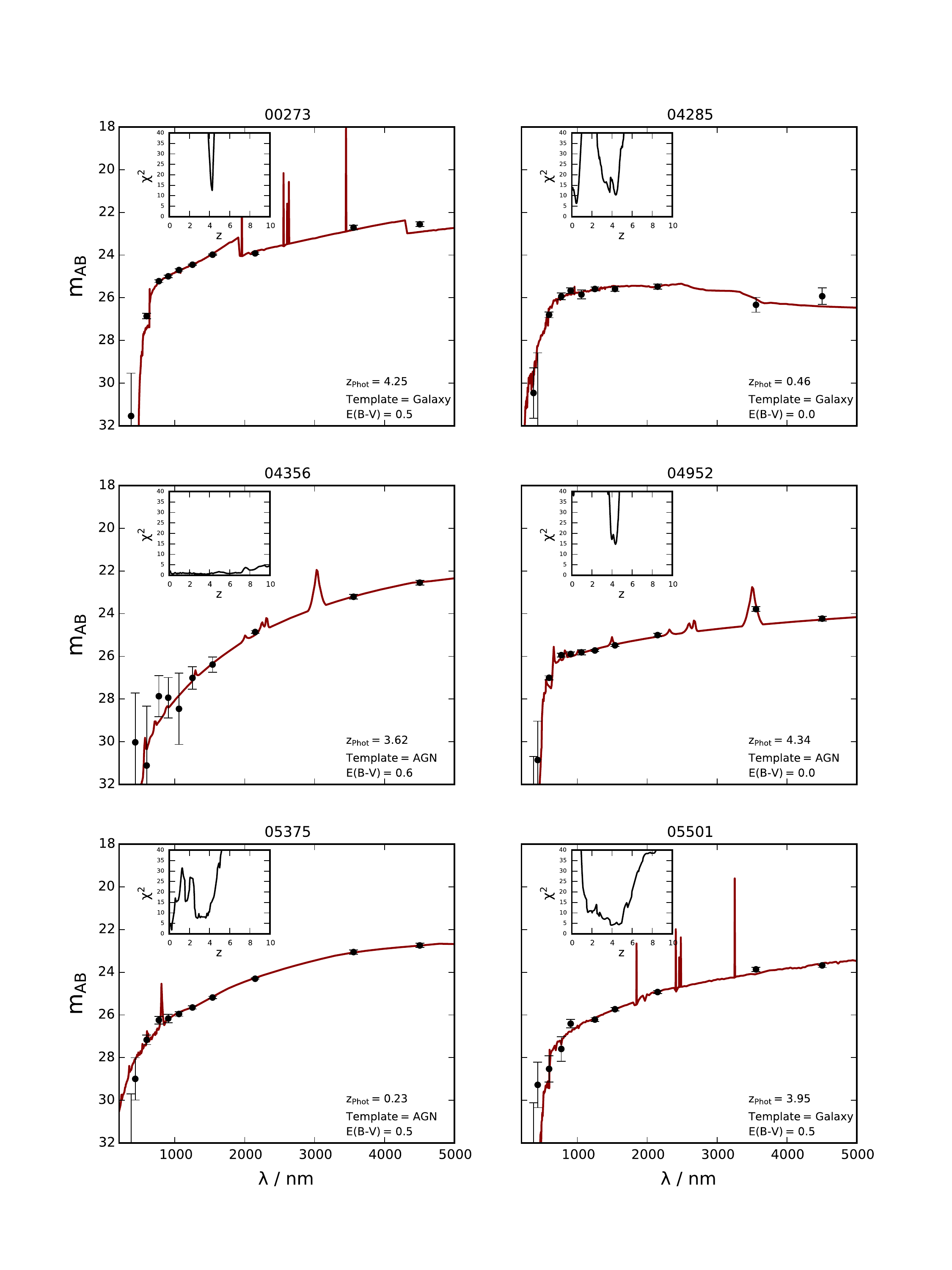}
\caption{The best-fitting spectral energy distributions and inserts showing $\chi^{2}$ versus photometric redshift for the 22 AGN candidates listed in Table 1. The best-fitting photometric redshift, the type of the best-fitting SED template (galaxy or AGN) and the preferred dust reddening for each source is written in each panel. The sources are ordered by the identification number from 
the GOODS-S CANDELS catalogue of Guo et al. (2013), as also adopted by Giallongo et al. (2015), which is written above each panel.}
\end{figure*}
\begin{figure*}
\includegraphics[width=\textwidth, height=650pt]{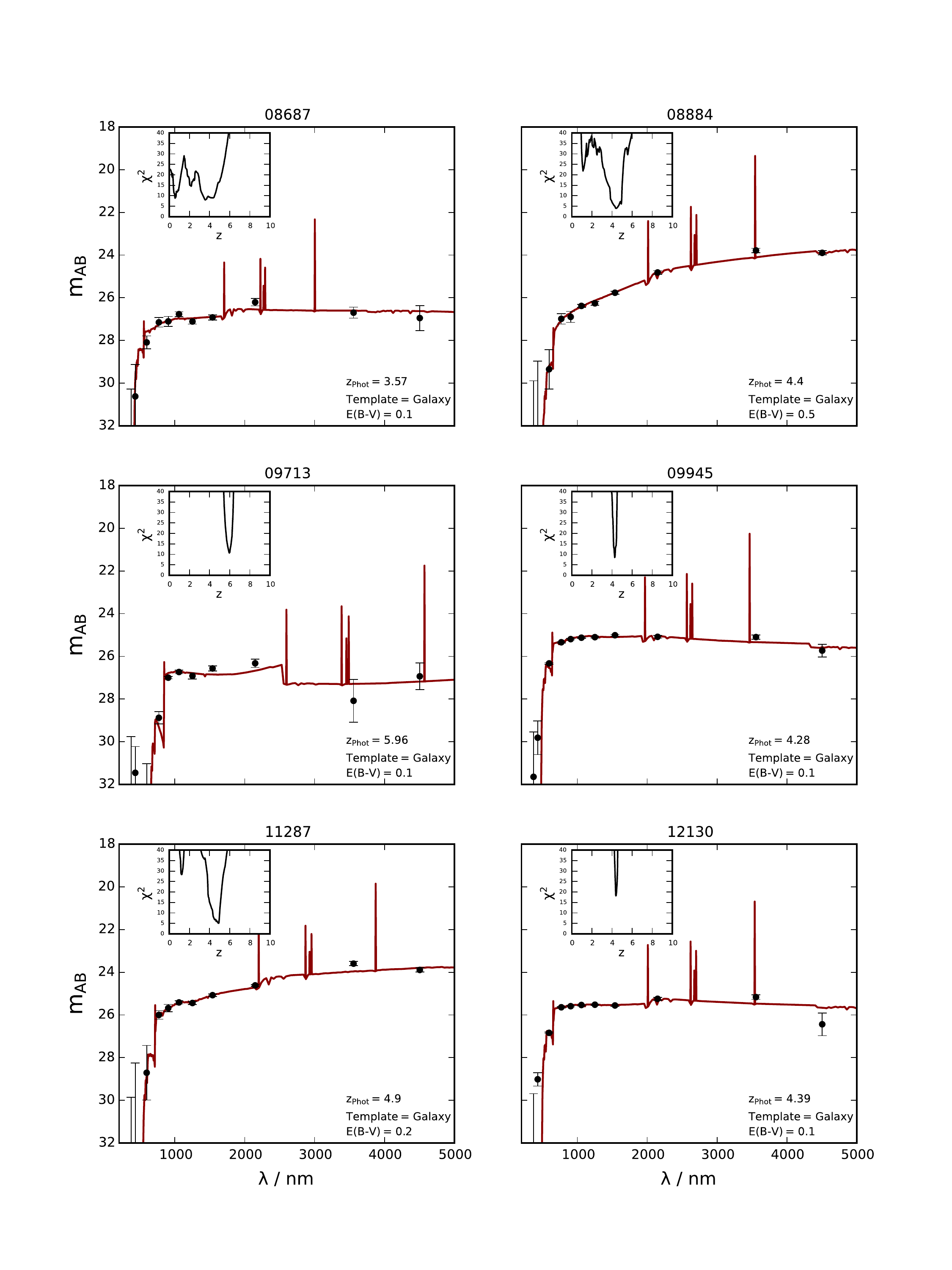}
\ContinuedFloat
\caption{(continued)}
\end{figure*}
\begin{figure*}
\includegraphics[width=\textwidth, height=650pt]{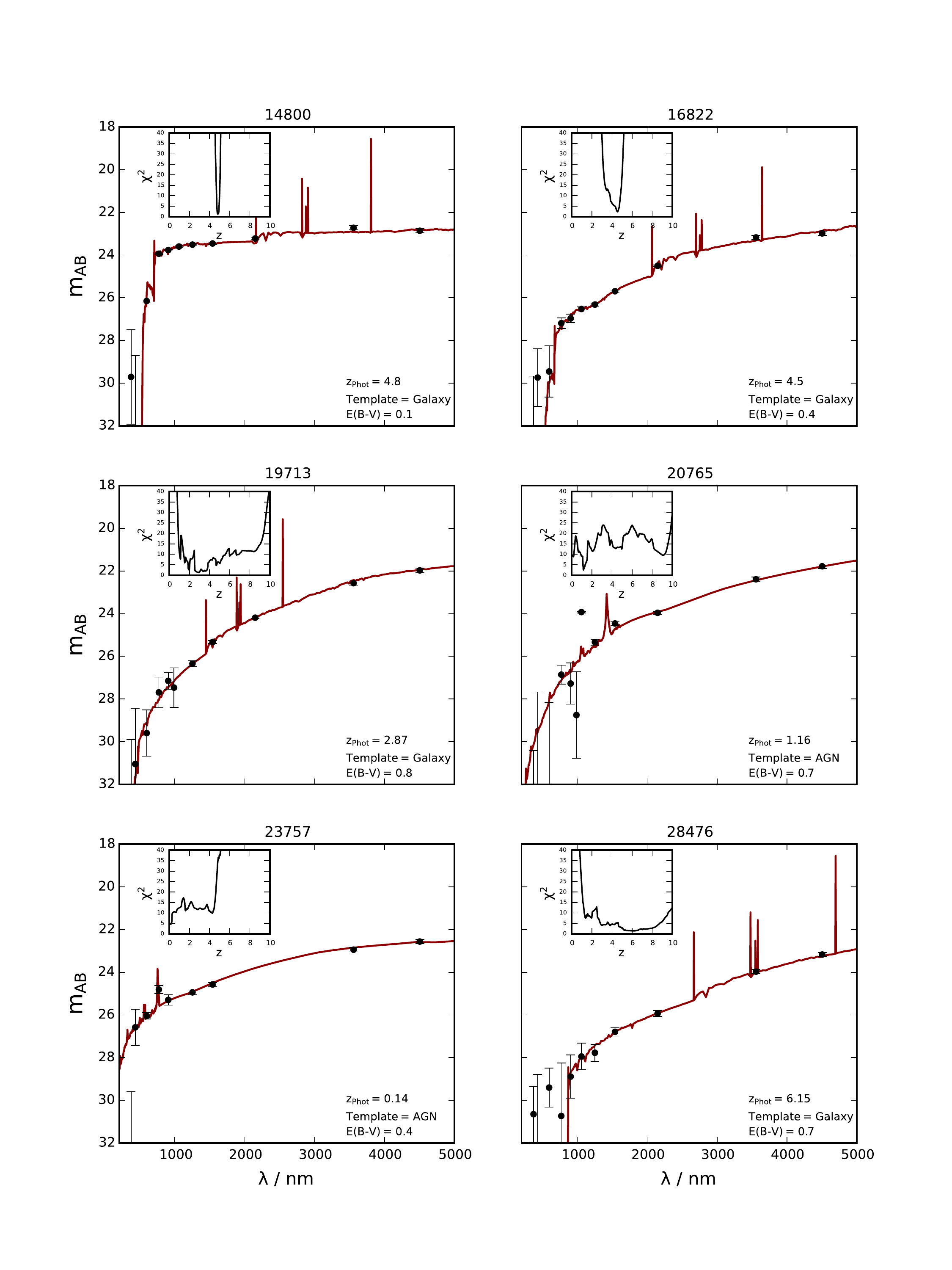}
\ContinuedFloat
\caption{(continued)}
\end{figure*}
\begin{figure*}
\includegraphics[width=\textwidth]{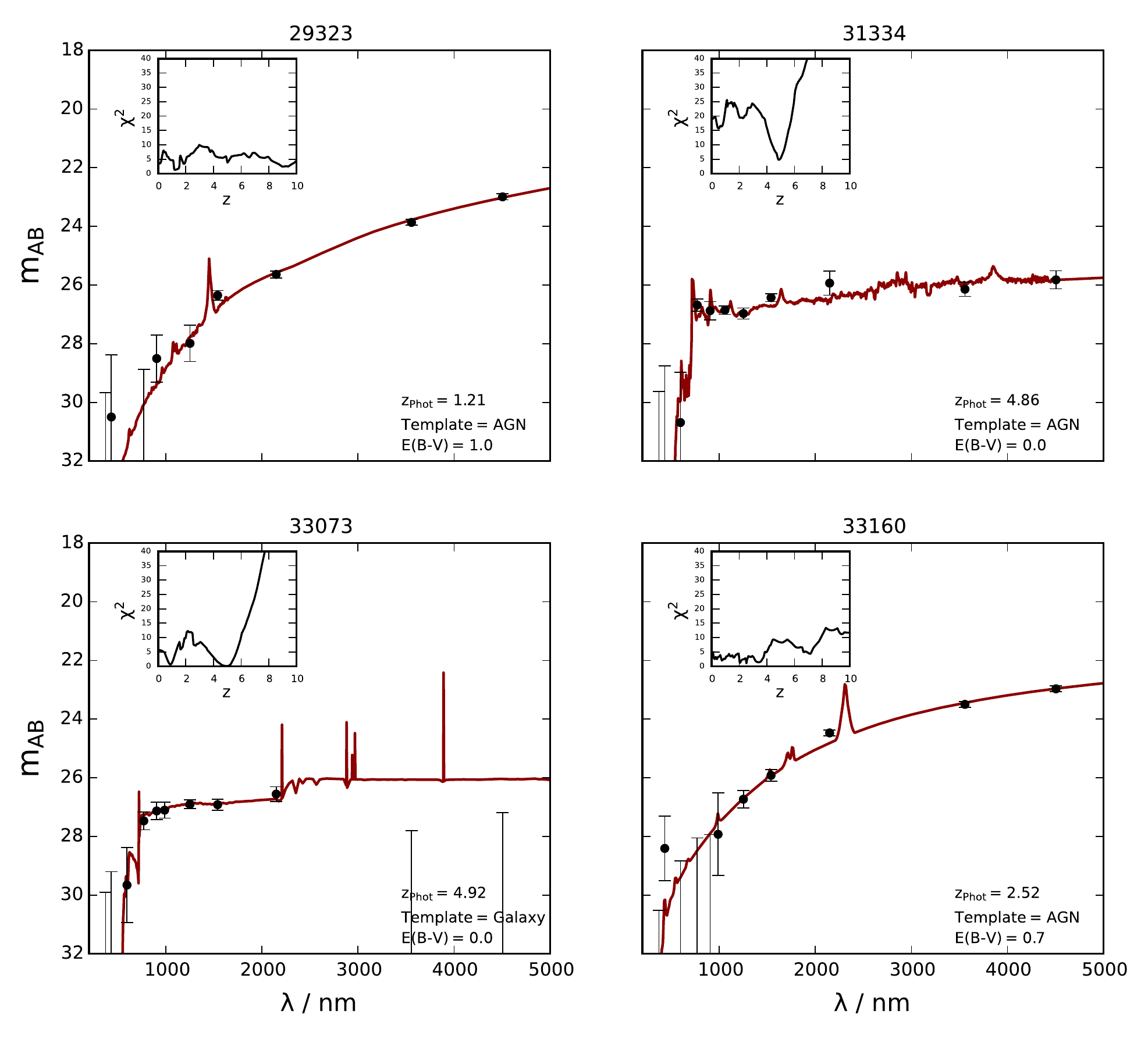}
\ContinuedFloat
\caption{(continued)}

\end{figure*}

\section{Reanalysis of the 22-source sample}

\subsection{Photometric redshifts}

A comparison of columns 7, 8 and 9 in Table\,1 shows that, whilst for many sources the various pre-existing photometric redshifts are in reasonable agreement, this is certainly
not the case for all of the objects. In particular, for several sources the photometric redshifts reported by Hsu et al. (2014) (who included AGN templates 
in their SED fitting) are substantially lower than those reported by Giallongo et al. (2015) or Skelton et al. (2014) (who considered only galaxy templates). The differences may be due to
the inclusion of different data (e.g. Hsu et al. included medium-band photometry), different template sets, different allowed ranges of reddening, or a different treatment of the photometric errors, but (unsurprisingly) the precise reasons for these differences are hard to isolate.

Therefore, given the importance of properly establishing the redshift distribution of these candidate high-redshift AGN, we decided to reinvestigate in detail the photometric redshifts of all 
22 claimed optical+near-IR AGN counterparts, building on the photometric redshift work performed in Parsa et al. (2016), utilising 
a mix of both galaxy and AGN templates, and (crucially) allowing for dust reddening extending up to $E(B-V) = 1.6$ (as demonstrated by Dunlop et al. (2007), allowing dust-reddening
to range up to values as high as $A_V \simeq 6$, or equivalently $E(B-V) \simeq 1.5$, is essential to avoid dusty intermediate-redshift galaxies being mis-classified as Lyman-break
galaxies at $z > 4$). 

As in Parsa et al. (2016), we used the public code Le Phare (PHotometric Analysis for Redshift Estimate; Ilbert et al. 2006) to determine the photometric redshifts. This utilises a standard template fitting technique, with the best-fitting spectral template and associated photometric redshift determined by comparing the template SEDs with the observed photometry via $\chi^{2}$ 
minimisation.\footnote{\url{http://www.cfht.hawaii.edu/~arnouts/LEPHARE/lephare.html}}

To ensure the correct treatment of non-detections, we performed the fitting in flux-density--redshift space (rather than magnitude--redshift space) and set realistic floors to 
the photometric uncertainties as described in Parsa et al. (2016). The increasing absorption of the IGM with redshift was modelled according to Madau (1995), and dust reddening 
was implemented using the Calzetti et al. (2000) attenuation law, with reddening allowed to vary over the range $0 < E(B-V) < 1.6$ (with steps of $\Delta{E(B-V)}=0.1$).

This implementation of the Le Phare code was applied to the sample of 22 sources in two-template mode, allowing for combinations of galaxy and AGN model SEDs. For the galaxy template set, we adopted PEGASEv2.0 (Fioc \& Rocca-Volmerange 1999) template SEDs with emission lines switched on over the redshift range $z= 0.1-10$, and, in addition, we adopted four sets of AGN model templates over the same redshift range. The best solution was then selected based on the lowest $\chi^{2}$ achieved over the complete galaxy+AGN template set, and the associated photometric redshift adopted. The best-fitting SEDs are shown in Fig.\, 1, and the corresponding preferred photometric redshifts are recorded in column 10 of Table\,1. As can be judged from the $\chi^2$ values of the best-fitting SEDs (column 11 of Table\,1), acceptable fits were achieved in all cases. The inserts in Fig.\,1 showing $\chi^2$ versus redshift can be used to judge the 
uncertainties in each photometric redshift.

Reassuringly, in the 5 cases where a robust spectrosopic redshift has already been determined (see column 6 in Table\,1) our photometric redshifts are in good agreement both with the spectroscopic values, and with the photometric redshifts derived in the other pre-existing studies (note that this does not, however, mean that we agree they are X-ray detected AGN; see below). 
However, for many other sources our new photometric redshifts are substantially lower than those reported by either Giallongo et al. (2015) or Skelton et al. (2014), and are often in better 
accord with (or even lower than) the values reported by Hsu et al. (2014). In the most dramatic examples of such discrepancies it appears that our best-fitting solution 
has been achieved with an AGN template (column 12 of Table\,1) but it is also the case that, for several sources, the preferred fit has a high level of dust reddening ($E(B-V)$; 
column 13 of Table\,1).

The differences between the resulting redshift distributions derived for this 22-source sample are illustrated in Fig.\,2.
 
\begin{figure}
\centering
\includegraphics[width=0.5\textwidth, height=530pt]{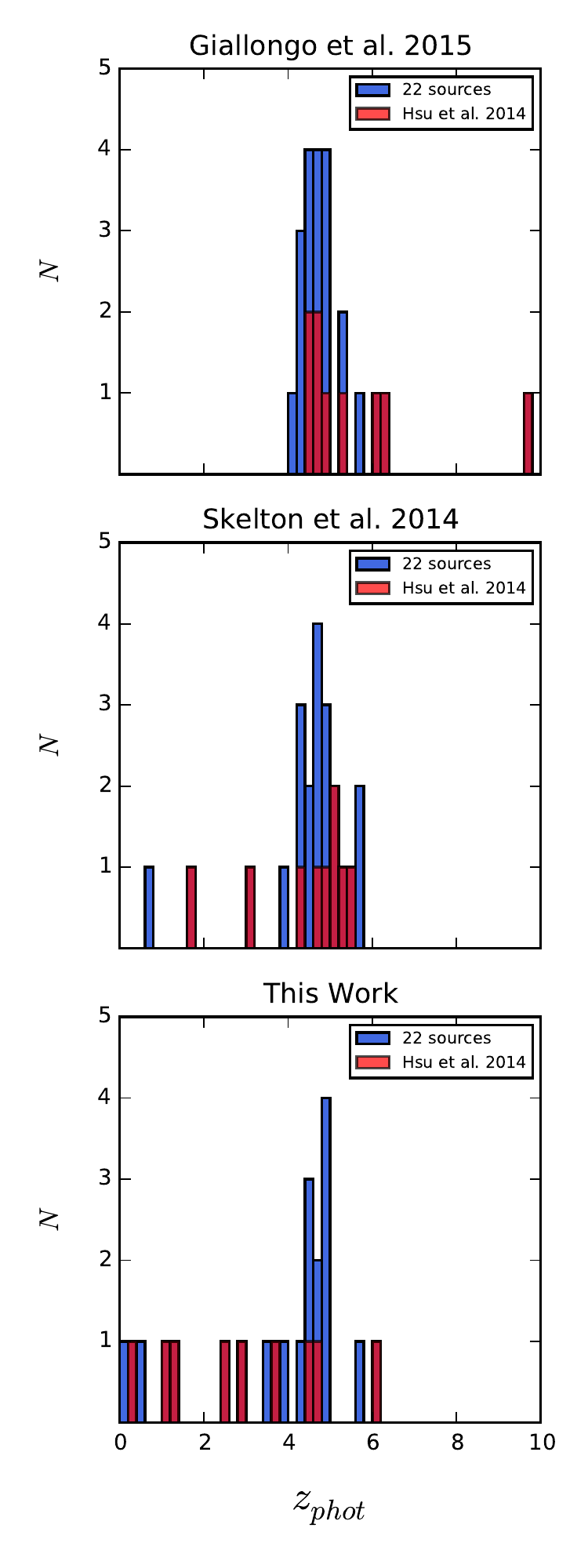}
\caption{Alternative redshift distributions for the 22 high-redshift X-ray AGN candidates reported by Giallongo et al. (2015) in the GOODS-S field. In all three panels, the 
5 sources with spectroscopic redshifts are plotted at the same (spectroscopic) values. In addition, in each panel we have indicated in red the subset of 8 brighter X-ray sources 
also studied by Xue et al. (2011) and Hsu et al. (2014), while the remaining subset of 14 fainter `sources' uncovered by Giallongo et al. (2014) (with the benefit of the $H$-band positional priors) 
is indicated by the blue histograms. The top panel 
shows the redshift distribution derived by Giallongo et al. (2015), where all 22 sources are found to lie at $z > 4$. The central panel shows the corresponding redshift information 
from the 3D-HST survey (Skelton et al. 2014). The bottom panel shows the results of our own photometric redshift analysis.}
\end{figure}

\subsection{Robustness of the X-ray detections}

We next carefully inspected the 4\,Ms {\it Chandra} X-ray imaging of each of the 22 sources. 
In Appendix A (Fig.\,A1) we show the 
soft, hard and full X-ray energy-band images (0.5--2 keV, 2--10 keV and 0.5--10 keV respectively) of each
source, after smoothing with a Gaussian function with $\sigma = 1.5$\,arcsec.

The 8 sources already found in the X-ray catalogue of Xue et al. (2011) are clearly detected in all three X-ray images. We also (possibly in some cases optimistically) confirm the detection of another 7 sources, at least in the soft X-ray imaging (ID: 8687, 8884, 9945, 11287, 14800, 23757, 28476). However, we are not able to confirm the detection of the remaining 7 candidates. Based on analysis of the raw or smoothed images, we can find no evidence 
of significant emission in the soft, hard or full bands at the optical/near-IR positions of these 7 sources (ID: 4285, 4952, 5501, 9713, 12130, 31334, 33073).

To explore this further we created stacks of the X-ray images. In Fig.\,3 we show the stacked soft, hard and full band images for the 15 X-ray sources for which we find at least some evidence for individual source detections.
Obviously the resulting stacked detection is highly-significant (it contains the 8 brightest X-ray sources, as well as the 7 more marginal detections), but this figure also provides reassurance 
of the accurate astrometric alignment between the X-ray and optical/near-IR positions that can be expected for genuine X-ray detections.
  
Then, in Fig.\,4 we show the corresponding stacked X-ray imaging of the 7 individually undetected sources.
Even after stacking, there is no evidence for any significant X-ray flux from this group of sources, casting 
further doubt on the reality of the X-ray detections claimed for these objects by Giallongo et al. (2015). 
It can of course be argued that the approach adopted here is somewhat cruder than the more 
sophisticated approach adopted in Giallongo et al. (2015) (which also incorporated the photon clustering
in time), but we still regard it as a matter of serious concern that a stack of these 7 sources shows no significant X-ray emission coincident with the optical/near-IR central position.

In summary, we find that while the approach of using $H_{160}$ prior positions to push the detection
limits of the {\it Chandra} imaging has probably successfully revealed an additional 7 sources which eluded
detection in the X-ray catalogue produced by Xue et al. (2011), a further 7 of the X-ray detections 
reported by Giallongo et al. (2015) appear to be erroneous (indeed, we predict these
7 sources will also remain undetected in the 7\,Ms {\it Chandra} imaging; Vito et al. 2016).

Combining this re-inspection of the X-ray data, and our re-analysis of the photometric redshifts, 
we can find evidence for only one possible X-ray source in GOODS-S at $z > 6$ (ID: 28476). 
However, even for this remaining high-redshift AGN candidate the SED fit shown in Fig.\,1 does not 
formally exclude the possibility of a significantly lower redshift solution. Indeed, this object
is arguably suprisingly bright to lie at $z > 6$, and in Fig.\,5 we show a zoom-in on the $\chi^2 - z$ 
plot for this source, which shows that, within $\Delta \chi^{2} = 4$ of the best-fitting solution, a 
redshift as low as $z \simeq 3$ cannot be excluded.

For the remainder of this paper we retain 
the $z \simeq 6.15$ redshift solution for this source, but the uncertainty over its redshift
means that our final estimate of AGN UV emissivity at $z \simeq 6$ should arguably be regarded 
as an upper limit.

\subsection{Completeness check}
The focus of this study is on revisiting the claimed 22-source $z > 4$ sample presented by
Giallongo et al. (2015). However, before proceeding to determine the implications of our revised
results for this sample, it is important to check whether our photometric redshift analysis within
GOODS-S yields any new X-ray detected sources at $z > 4$ which were {\it not} reported by Giallongo
et al. (2015).

Considering all sources in GOODS-S with $H_{160} < 27$, we find that there are 455 objects which Dahlen et al. (2013)
found to lie at $z > 4$ that we find lie at $z < 4$. However, we also found 95 objects
for which our photometric redshift lies at $z > 4$, while the Dahlen et al. (2013) results indicate $z < 4$. So, while it is true that,
in general, our photometric redshift technique (allowing higher reddening) yields many fewer high-redshift objects than
inferred from the Dahlen et al. (2013) results, there are inevitably cases where the opposite is true.

We inspected the X-ray imaging for all 95 of these objects, and found one detection, corresponding to the GOODS-S
source ID: 09035. For this source, we derive $z_{\rm phot} = 4.25$, whereas Dahlen et al. (2013) report $z_{\rm phot} = 0.64$.
We have therefore included this object in our analysis of the luminosity function and resulting luminosity density
as presented in the next section (with our inferred redshift, this object has an absolute UV magnitude of $M_{1500} = -20.45$).

\begin{figure*}
\centering
\includegraphics[scale=0.23]{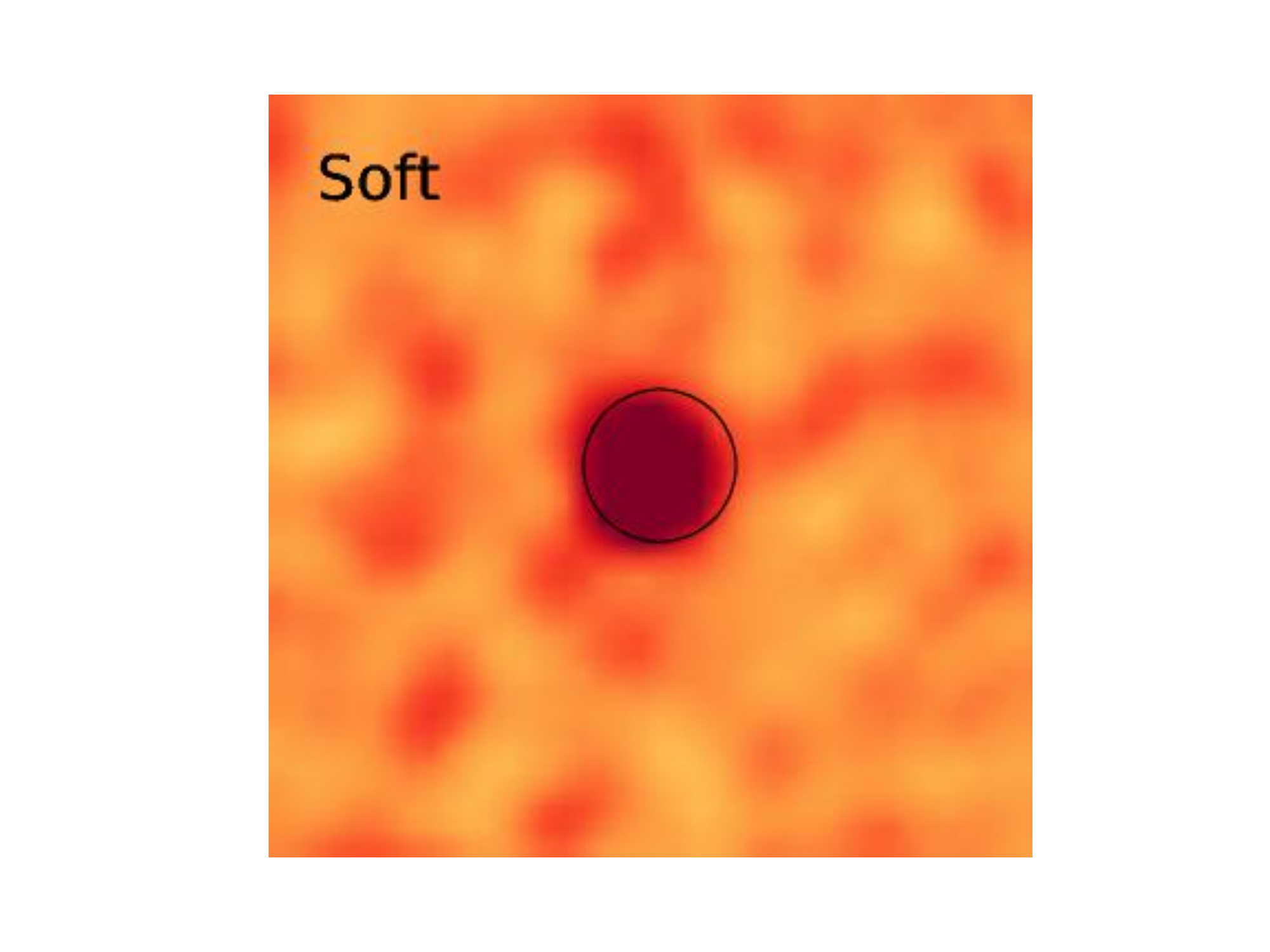}
\includegraphics[scale=0.23]{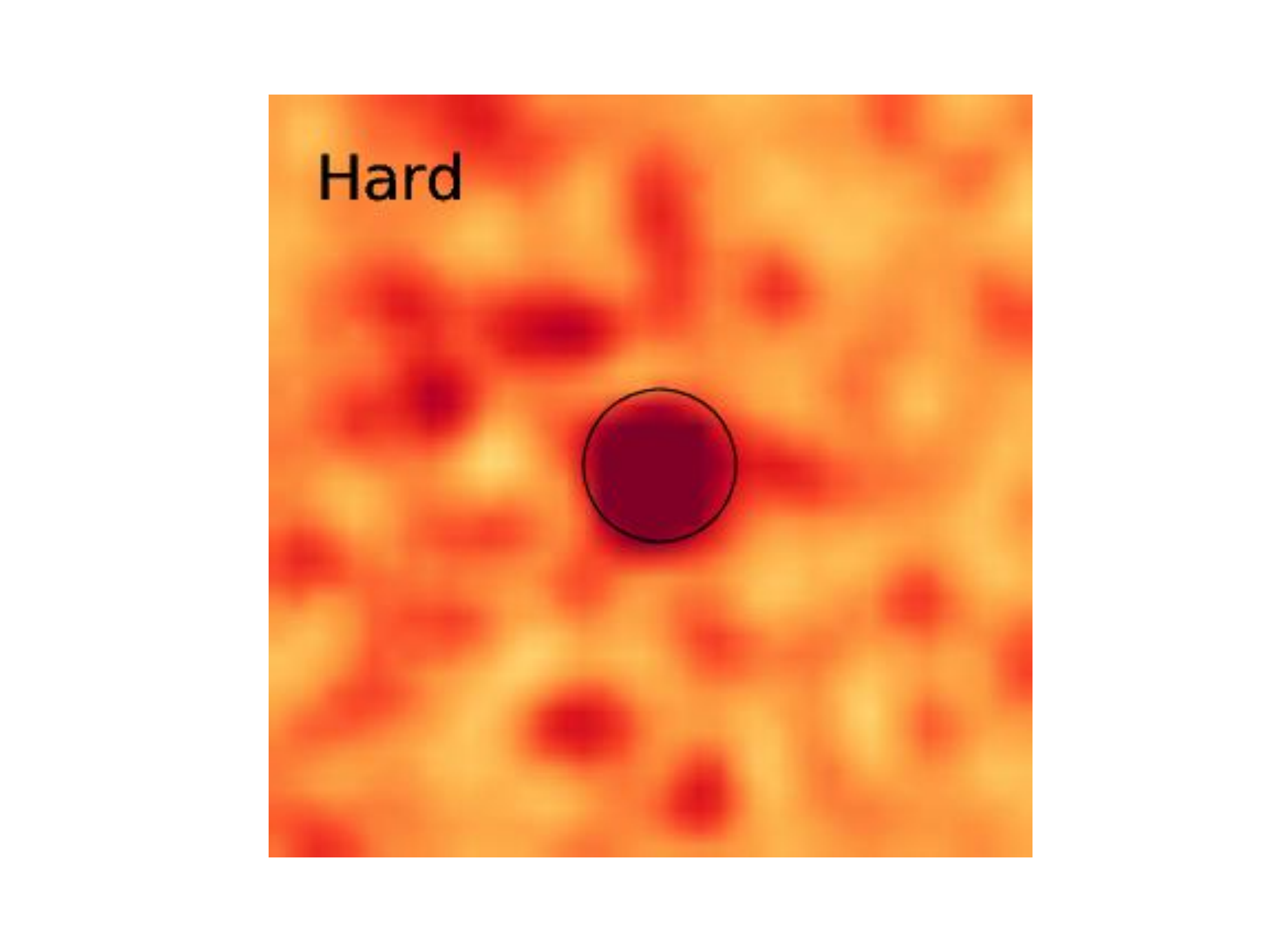}
\includegraphics[scale=0.23]{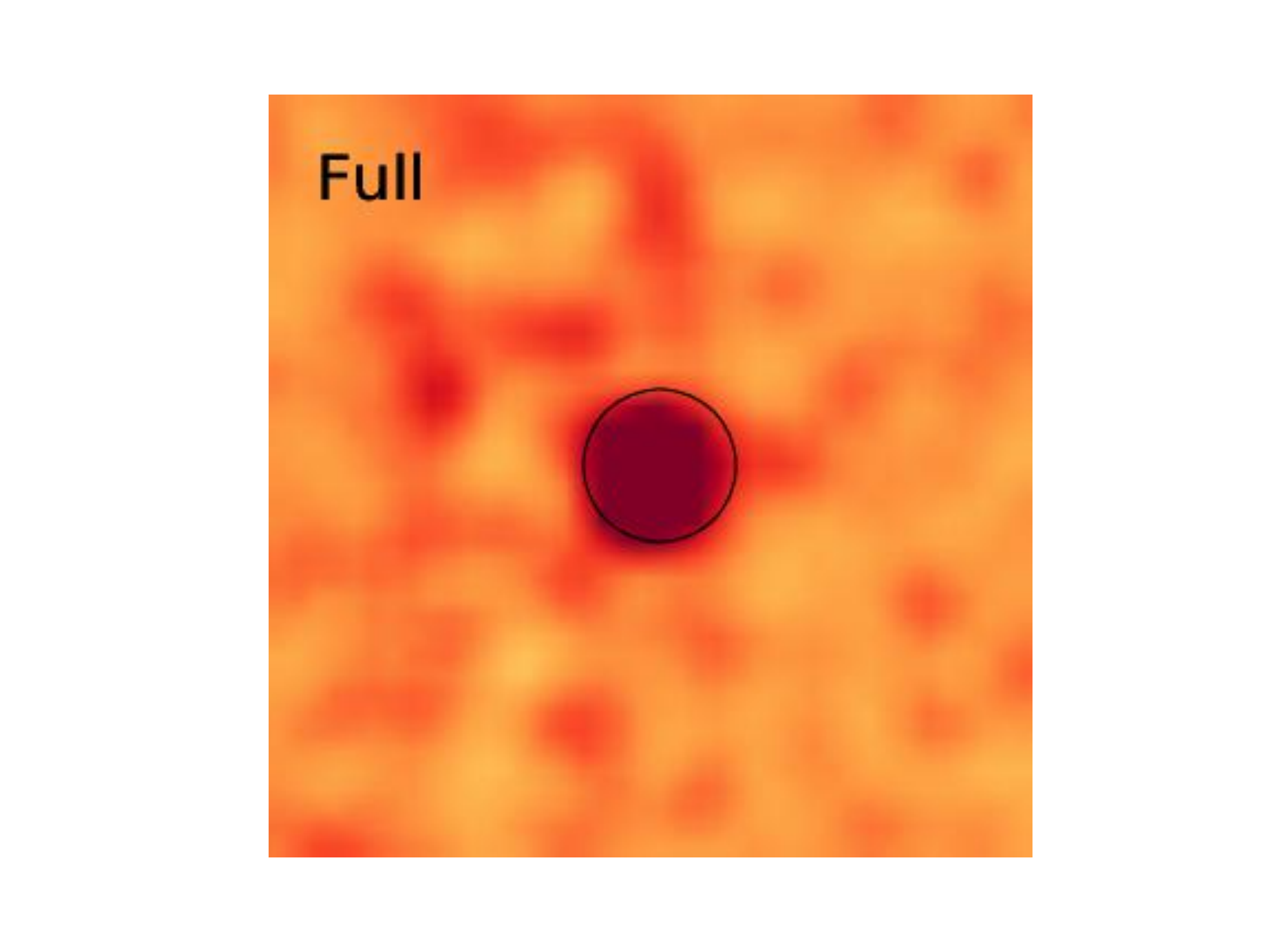}
\caption{Postage-stamp soft, hard and full band {\it Chandra} 4\,Ms X-ray images of the stack of the 
15 individually-detected sources in the 22-source sample. These stacks include the 
8 X-ray sources which already featured in the catalogues produced from the 4\,Ms imaging 
by Xue et al. (2011), plus the 7 more marginal detections confirmed in Fig.\,A1.
The imaging has been Gaussian smoothed with $\sigma = 1.5$\,arcsec, and each stamp is 20 x 20 arcsec in size, with North to the top and East to the left. The colour scale spans a flux-density range 
corresponding to $\pm2.5\sigma$ around the median background. 
The position of the {\it HST} WFC3/IR optical/near-IR counterpart is marked by a circle of radius 2\,arcsec.}
\end{figure*}

\begin{figure*}
\centering
\includegraphics[scale=0.23]{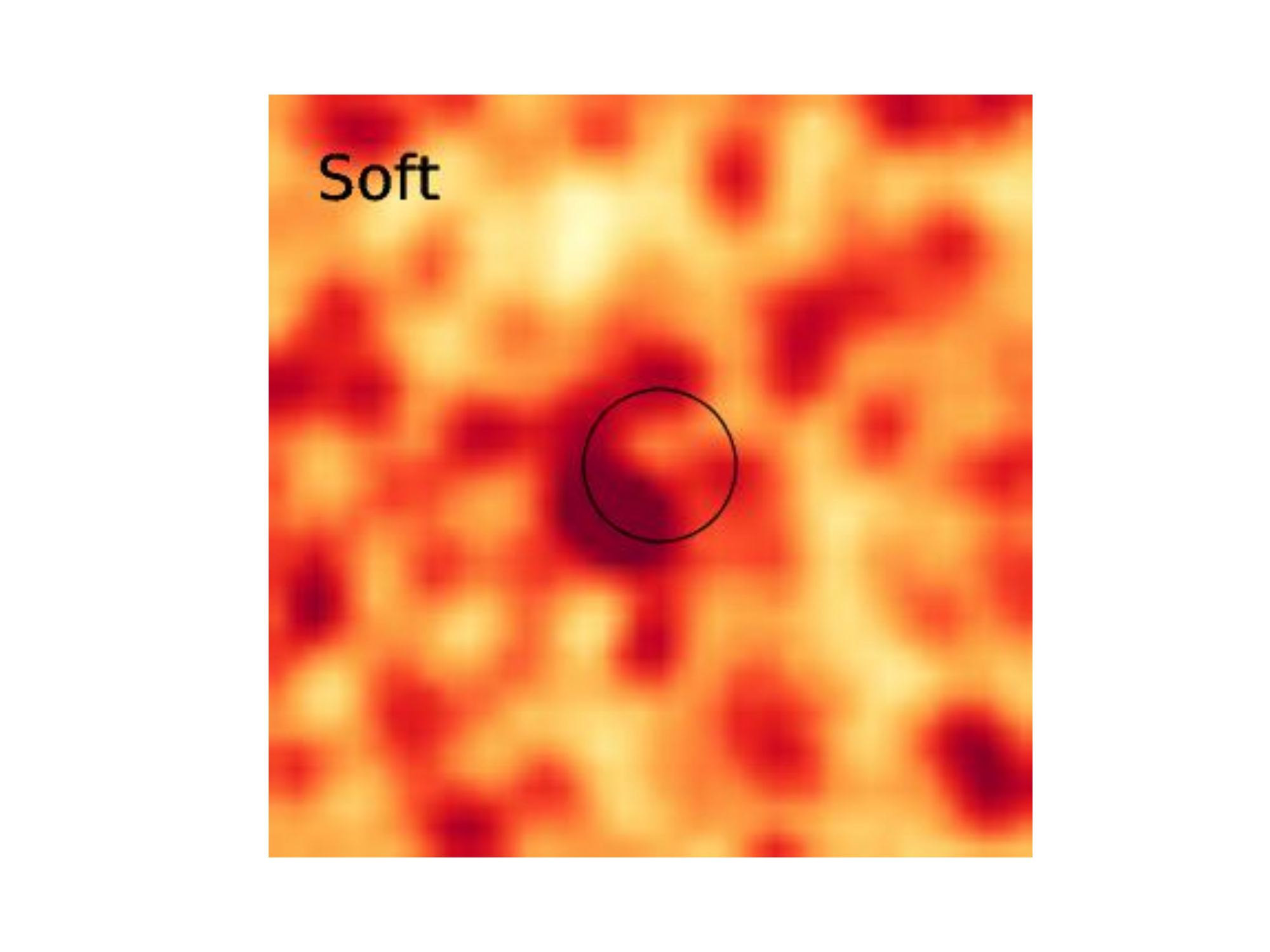}
\includegraphics[scale=0.23]{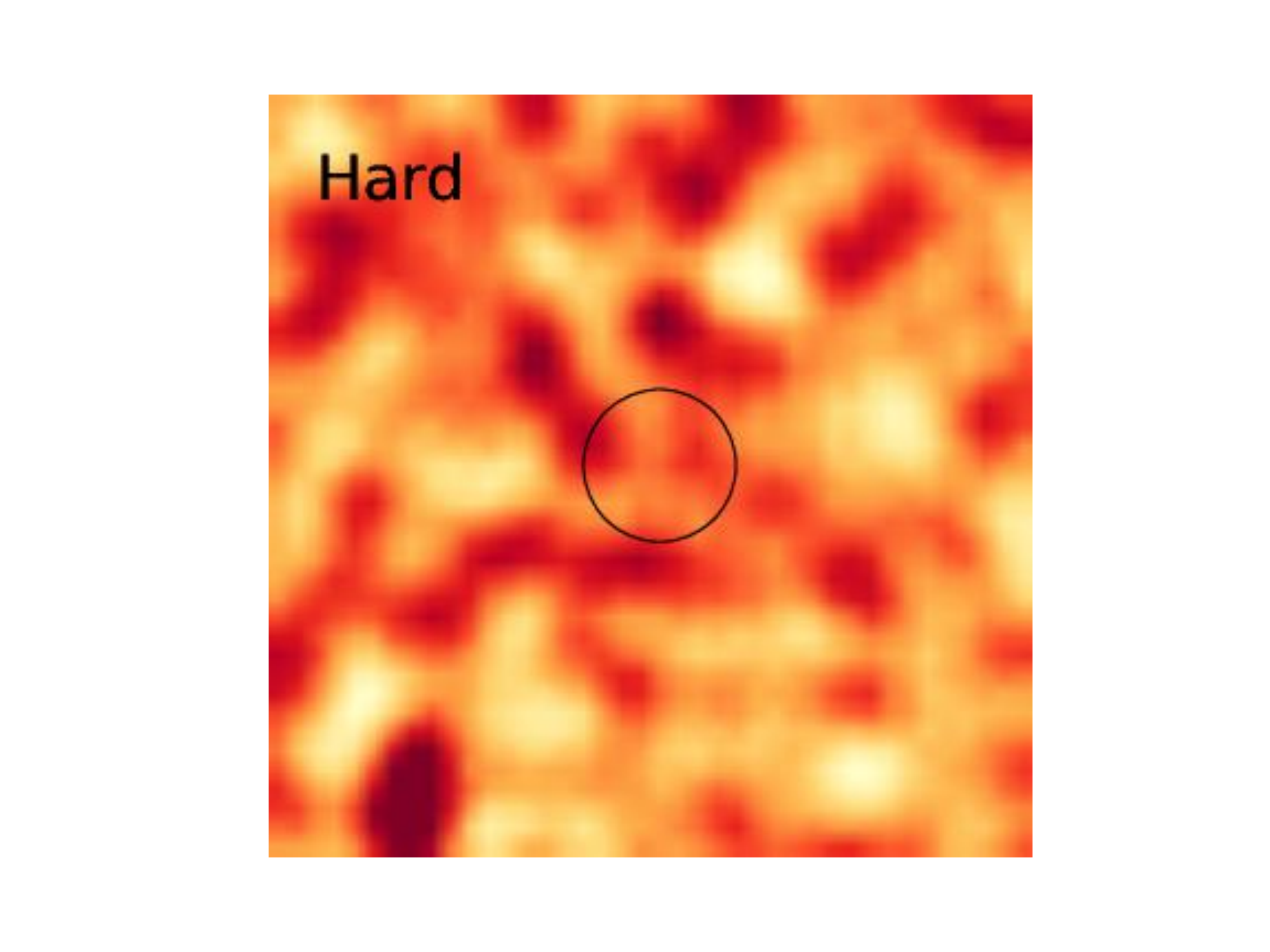}
\includegraphics[scale=0.23]{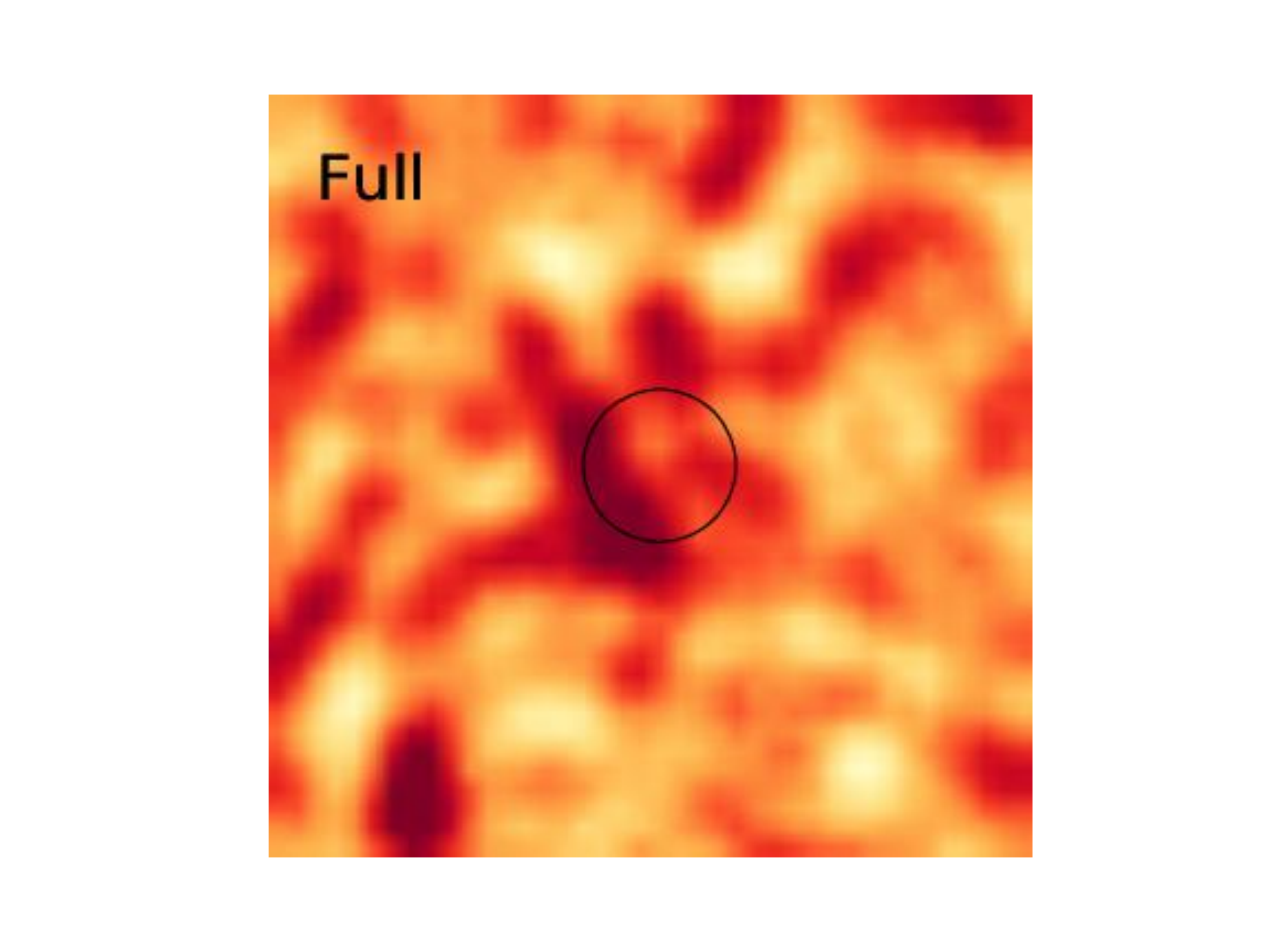}
\caption{Postage-stamp soft, hard and full band {\it Chandra} 4\,Ms X-ray images of the stack of the 
7 individually {\it undetected} sources in the 22-source sample (ID: 4285, 4952, 5501, 9713, 12130, 31334, 33073; see Appendix A, Fig.\,A1). 
As in Fig.\,3, the imaging has been Gaussian smoothed with $\sigma = 1.5$\,arcsec, and each stamp is 20 x 20 arcsec in size, with North to the top and East to the left. The colour scale spans a flux-density range 
corresponding to $\pm2.5\sigma$ around the median background. 
The position of the {\it HST} WFC3/IR optical/near-IR counterpart is marked by a circle of radius 2\,arcsec. It can be seen that, even in the stacked imaging, there is no evidence for significant emission coincident
with the central {\it HST} positions used as as positional priors by Giallongo et al. (2015).}
\end{figure*}

\section{Revised emissivity estimates}

\subsection{Calculating AGN hydrogen ionizing 
emissivity}

Since our focus here is on revisiting the results of Giallongo et al. (2015), we follow their approach 
to calculating the ionising emissivity of the high-redshift AGN population. 

Giallongo et al. (2015) applied the standard $1/V_{max}$ approach (Schmidt 1968) to determine the 
rest-frame Far-UV (1450\,\AA) LF of the X-ray AGN by binning their proposed sample of 22
$z > 4$ sources into three redshift bins corresponding to $z =$ 4.25, 4.75 and 5.75. 
Absolute magnitudes, $M_{1450}$, were computed from the apparent magnitudes in the filters 
closest to sampling $\lambda_{rest} = 1450$\,\AA\ within each redshift bin. Using this technique, they 
reported new measurements of the far-UV LF for AGN down to a faint limit corresponding to
$M_{1450} \simeq -18.5$ over the redshift range $z = 4 - 6.5$. 

After deriving the shape of LF at these three redshifts, Giallongo et al. (2015) then 
calculated the inferred hydrogen ionizing emissivity of these AGNs using:

\begin{equation}\begin{split}
& \epsilon_{ion} (z)=\langle f\rangle \epsilon_{912}=\\
&\langle f\rangle \int \phi(L_{1450},z)L_{1450}\left({1200\over 1450}\right)^{0.44}\left({912\over 1200}\right)^{1.57}dL_{1450} 
\end{split}\end{equation}
where  $\langle f\rangle$ is the average escape fraction of ionizing radiation from the AGN host galaxies and $\epsilon _{912}$ is the co-moving emissivity produced by the volume-averaged AGN activity. The shape of the AGN SED from $\lambda=1450$\,\AA\ to $\lambda=912$\,\AA\ was represented by a double power-law with slopes adopted following Schirber et al. (2003) and Telfer et al. (2002). 

In applying the above formula, Giallongo et al. (2015) 
calculated the integral down to a faint-luminosity limit corresponding to $M_{1450,limit}=-18$, and 
assumed an average escape fraction of $\langle f\rangle$ = 1.


\begin{figure}
\centering
\includegraphics[width=0.5\textwidth]{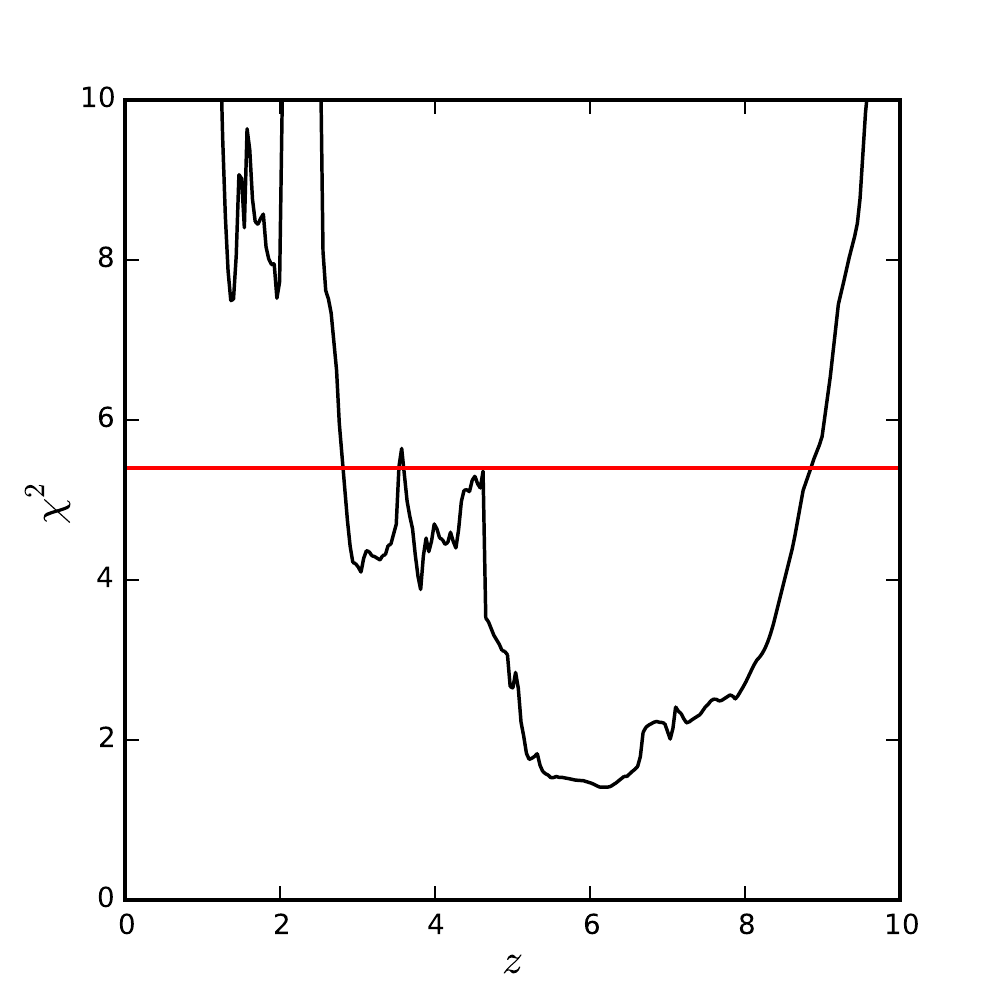}
\caption{The range of possible redshift solutions for the one potential remaining 
$z \simeq 6$ X-ray detected AGN in our revised sample (ID-28476). The horizontal red line 
indicates $\Delta\chi^2=4$ above the minimum $\chi^2$ achieved for the formal best solution at 
$z \simeq 6$, showing that photometric redshifts as low as $z \simeq 3$ are acceptable, even without
application of any luminosity prior (which would favour lower redshifts). We note that Giallongo et al. (2015)
also found that this source had a highly-uncertain photometric redshift.}
\end{figure}


\begin{table*}
\caption{The high-redshift co-moving hydrogen ionizing emissivity provided by AGN based on a 
a simple rescaling of the number of confirmed X-ray sources in each redshift bin.}

\begin{tabular}{l|c|c|c|c}
\hline
 \multirow{2}{*}{Redshift Bin}& \multicolumn{2}{c|}{Giallongo et al. 2015} & \multicolumn{2}{c}{This Work}\\
\cline{2-5}
&$\epsilon_{912} / 10^{24}\, {\rm erg\,s^{-1} Hz^{-1} Mpc^{-3}}$ & N$_{obj}$ &  $\epsilon_{912} / 10^{24}\, {\rm erg\,s^{-1} Hz^{-1} Mpc^{-3}}$ & N$_{obj }$\\
\hline
  4.0 - 4.5 (4.25) & 6.80 & 6  & 3.40 & 3 \\
  4.5 - 5.0  (4.75) & 5.92 & 8 & 2.20 & 3 \\
  5.0 - 6.5 (5.75) & 2.50 & 5 & 0.50 & 1 \\
\hline
\end{tabular}

\end{table*}


\subsection{Emissivity estimates from re-scaling of the high-redshift AGN number counts}

We first perform the basic calculation of simply rescaling the emissivities reported by Giallongo et al. (2015) according to the number of GOODS-S X-ray 
AGN that we now find lie in each of the redshift bins considered (i.e. at redshifts $z = 4.25$, 5.25 and 5.75). The numbers in each bin change both because we can confirm
only 15 of the original 22 sources as X-ray emitters, and because of the revised photometric redshifts given in Table\,1 and Fig.\,2.

The results are compared in Table\,2, where it can be seen that the estimates are not only reduced at all three redshifts, but progressively so with increasing redshift
(because we find only 50\% of the original number of X-ray sources in the redshift range $z = 4 - 4.5$,  only 38\% at $z = 4.5 - 5.0$, and only 20\% at $z = 5.0 - 6.5$).
Note that only one X-ray source remains in our sample at $z > 5.0$, approaching the reionization epoch. Moreover, as discussed above, even this source (ID:\,28476) could in fact lie at 
a redshift as low as $z \simeq 3$ (see Fig.\,5).

The revised ionizing emissivity estimates given in Table~2 are plotted and discussed further in Section 5.


\begin{table*}
\caption{AGN luminosity functions and resulting derived ionizing emissivities. Binned number densities derived from the $V_{max}$ method are 
given for the luminosity bins which contain GOODS-S objects in our revised X-ray AGN sample, and best-fitting parameter values are given for the double power-law parameterisation of 
the far-UV LF given by Equation 2. 
The resulting derived ionizing emissivities, $\epsilon_{912}$, given in the final column have been calculated from the LFs using Equation 1, integrating down to 
$M_{1450} = -18$. As in Table 2, $\epsilon_{912}$ is given in units of $10^{24}\, {\rm erg\,s^{-1} Hz^{-1} Mpc^{-3}}$. Note that the
single object in the $4.0 < z < 4.5$, $M_{1500} = -20$ bin is the single extra source reported in Section 3.3.}
\begin{tabular}{c| c c c c  c c c  c }
\hline\hline
$\Delta z$ & $M_{1500}$ & $\log \phi_{obs}$ &  N$_{obj}$ & $\beta$ & $\gamma$ & $M^*$ & $ \log \phi^*$ & $ \epsilon_{912}$ \\
\hline\hline
$4.0-4.5$ & & & & $1.75\pm0.91$ & $2.74\pm0.14$ & $-23.37\pm1.25$ & $-5.82\pm0.72$ & 3.09\\ 
 & $-19$ & $-3.99\pm0.50$ & 1 & & & & &  \\ 
 & $-20$ & $-5.07\pm0.56$ & 1 & & & & & \\
 & $-21$ & $-5.20\pm0.56$ & 2 & & & & &  \\ 
\hline
$4.5-5.0$ & & & & $1.75$\,(fixed) & $2.56\pm0.24$ & $-23.10\pm4.02$ & $-6.13\pm2.08$ & 1.25  \\
 & $-19$ & $-$ & 0 & & & & &  \\
 & $-20$ & $-$  & 0 & & & & &  \\
 & $-21$ & $-4.80\pm 0.40$ & 2 & & & & &  \\
 & $-22.5$ & $-5.30\pm 0.70$ & 1 & & & & & \\
\hline
$5.0-6.5$ & & & & $1.75$\,(fixed) & $2.41\pm0.40$ & $-22.58\pm4.88$ & $-6.62\pm2.21$ & 0.26 \\
 & $-19$ & $-$ & 0 & & & & &  \\
 & $-20$ & $-5.99\pm0.66$ & 1 & & & & & \\
 & $-21$ & $-$ &  0 & & & & & \\
\hline
\end{tabular}\\
 
\end{table*}

\subsection{Emissivity estimates from a new determination of the evolving AGN LF}

The simple calculation presented above helps to clarify the extent to which the proposed high-redshift sample of X-ray AGN is altered by the reinvestigation presented here.
However, a more careful reassessment of the resulting impact on inferred ionizing emissivity requires a redetermination of the far-UV LF to be utilised in Equation 1.


\begin{figure*}
\centering
\includegraphics[width=0.8\textwidth]{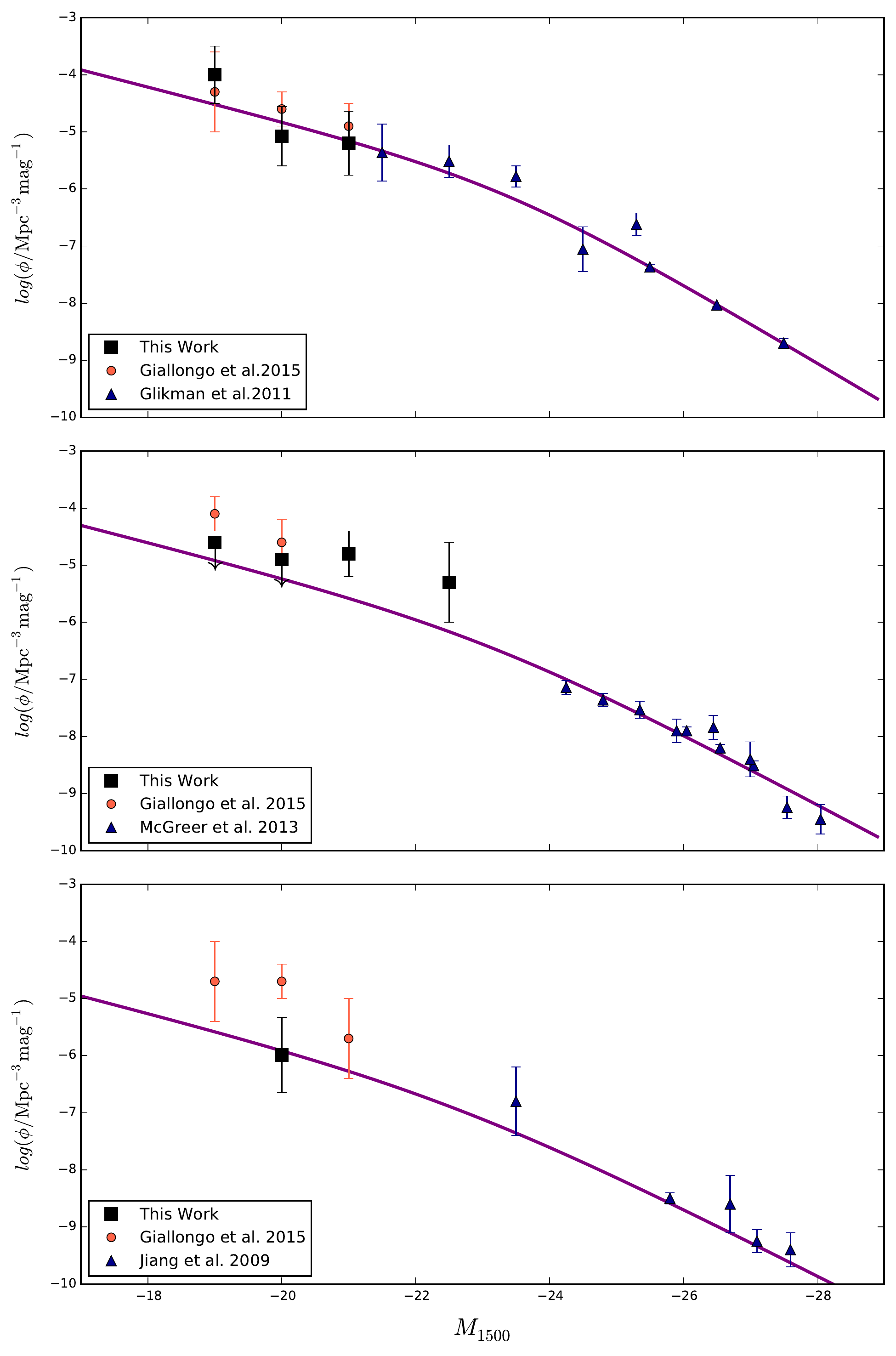}
\caption{The far-UV AGN luminosity functions at $z = 4.25$, 4.75 and 5.75, based on the revised $z > 4$ X-ray AGN sample produced in the present study, 
and the results from various brighter quasar surveys (as indicated in the legend in each panel). The small red circles and large black squares represent the results computed by 
Giallongo et al. (2015) and in this work respectively. The double power-law LF fitted at $z \simeq 4.25$ here is very similar to that derived by Giallongo et al. (2015), and essentially 
identical to that derived by by Manti et al. (2017) (with a faint-end slope $\beta \simeq 1.75$). In the two higher redshift bins, the lack of faint AGN results in a steady reduction in $\phi^*$.}
\end{figure*}


We therefore fitted a double power-law LF, parameterised as:

\begin{equation}
\phi=\frac{\phi^*}{10^{0.4(M^*-M)(\beta-1)}+10^{0.4(M^*-M)(\gamma-1)} }
\end{equation}

\noindent
where $\phi^{*}$, $M^{*}$, $\beta$ and $\gamma$ are the normalization, the break magnitude, the faint-end slope and the bright-end slope respectively.

We fitted this parametric LF to our revised GOODS-S comoving number densities along with the results at brighter magnitudes 
from the studies of Jiang et al. (2009), Glikman et al. (2011) and McGreer et al. (2013). Again, we attempted to calculate the far-UV LF for the three redshift bins corresponding 
to $z = 4.25$, 4.75 and 5.75.

The results of this fitting are shown in Fig.\,6, where, for the faintest three bins ($M_{1500}= -21 , -20, -19$) we show our new values along with the values calculated by Giallongo 
et al. (2015) for ease of comparison. We performed a completely free fit of the double power-law LF to the data in the $z = 4.25$ bin, because within this bin the data are of sufficient
quality to enable a meaningful estimate of the faint-end slope, $\beta=1.75\pm0.91$. We then fixed the faint-end slope at this value for the two higher-redshift bins, and fitted 
for $M^{*}$ and $\phi^*$. The resulting fits are shown by the curves in Fig.\,6, and the best-fitting parameter values are given in Table\,3.

The double power-law LF fitted at $z \simeq 4.25$, with a faint-end slope $\beta \simeq 1.75$, is very similar to that derived by Giallongo et al. (2015) (who actually report 
a slightly flatter faint-end slope), and indeed essentially 
identical to that recently derived by Manti et al. (2017). In the two higher redshift bins, the relative lack of faint AGN in our revised sample 
results in a steady reduction in $\phi^*$, and a mild monotonic reduction in the break luminosity. In contrast, Giallongo et al. (2015) reported essentially no evolution 
between $z \simeq 4.75$ and $z \simeq 5.75$. 

To calculate the the ionizing emissivities inferred by these far-UV LFs we follow Giallongo et al. (2015), and use Equation 1, with the same double power-law SED to extrapolate 
from 1450\,\AA\ to 912\,\AA, again integrate down to $M_{1450} = -18$, and assume an average ionizing photon escape fraction of $\langle f\rangle = 1$. The resulting emissivity estimates are given in the final column of Table\,3, and are even lower than the values 
estimated in Table\,2. These values are plotted and discussed further in the next Section.

\section{Discussion}

\begin{figure*}
\centering
\includegraphics[width=0.6\textwidth]{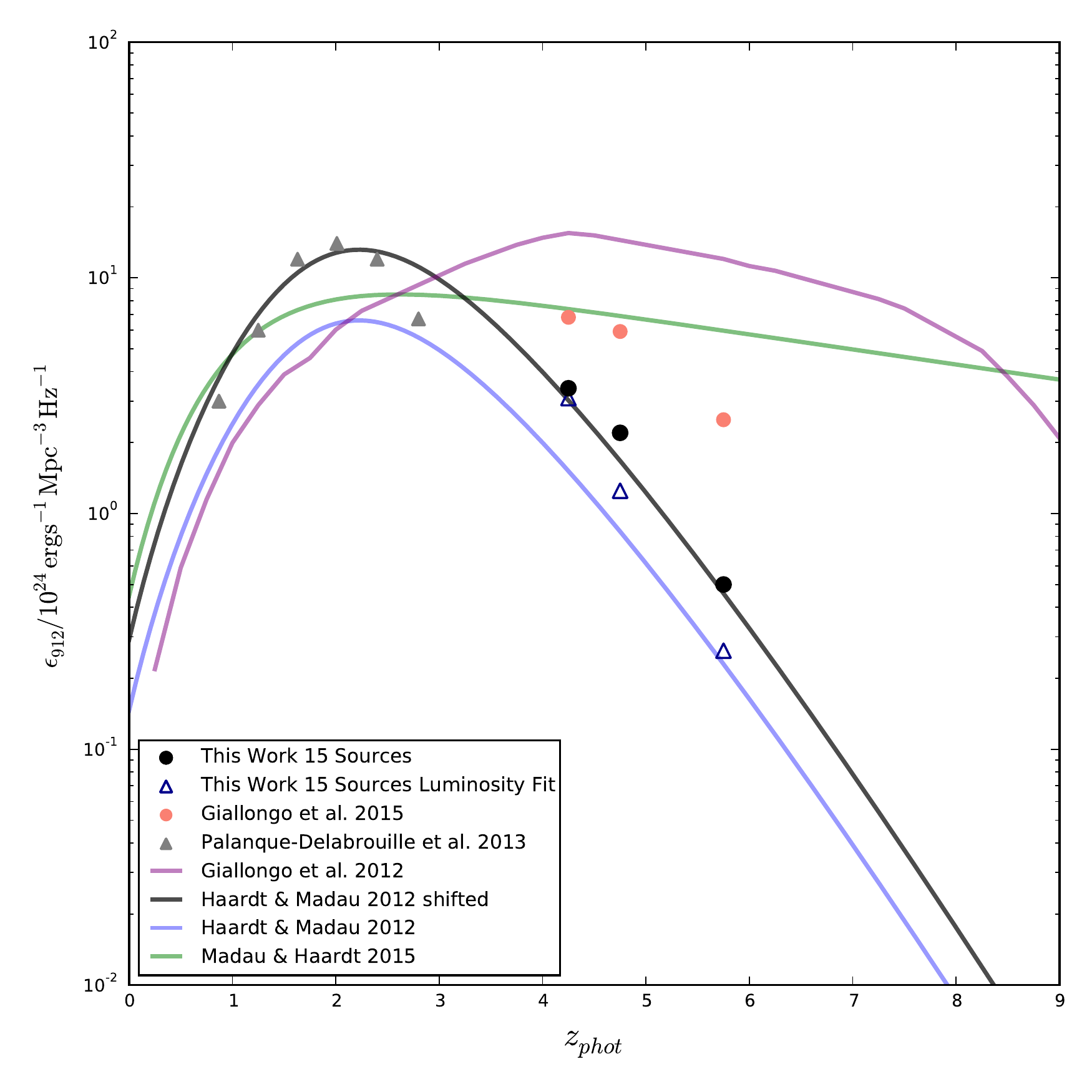}
\caption{Cosmic ionizing emissivity density, $\epsilon_{912}$, contributed by the evolving AGN population as a function of redshift. 
The pink circles show the emissivities calculated by Giallongo et al. 
(2015). The black filled circles and open triangles indicate our revised results, the former based on simple rescaling of the number of confirmed sources in each 
redshift bin (Table\,2), and the latter based on the luminosity-weighted integration of our revised far-UV LFs (Table\,3 and Fig.\,6). At $z < 4$ we plot the results from Palanque-Delabrouille et al. (2013). The purple line is from the model of Giallongo et al. (2012), the green line is from the model of Madau \& Haardt (2015), while the blue line is from the earlier 
model of Haardt \& Madau (2012). Our revised results at $z > 4$, and the latest results at lower redshifts, 
are more consistent with the redshift evolution described by the original model of Haardt \& Madau (2012). Increasing the amplitude of this model by a factor $\simeq 2$ (black line, and equation 3) provides 
a good description of the observed evolution of $\epsilon_{912}$.}
\end{figure*}



Our revised AGN hydrogen ionizing emissivity results at $z > 4$ are plotted in Fig.\,7, along with the results from Giallongo et al. (2015), values derived at lower redshifts 
by Palanque-Delabrouille et al. (2013), and the predictions of the studies of 
Giallongo et al. (2012), Haardt \& Madau (2012) and Madau \& Haardt (2015). The observational results of Giallongo et al. (2015) are in reasonable accord with the predictions of the 
theoretical modelling undertaken by Giallongo et al. (2012), which suggested that AGN could indeed make a significant contribution to cosmic hydrogen reionization. However, it can be seen 
that our revised results, either from the revised number counts (Table\,2) or from the luminosity-weighted integration of the revised LFs (Table\,3), are in much better accord with the predictions of the parametric model of Haardt \& Madau (2012), in which comoving ionizing emissivity of the AGN population peaks at $z \simeq 2$ and then declines rapidly at higher redshift. Indeed 
our revised emissivity estimates at $z > 4$, combined with the results derived from the quasar LFs produced by Palanque-Delabrouille et al. (2013) at lower redshifts, 
are very well described by the model of Haardt \& Madau (2012), simply boosted in amplitude by a factor of $\simeq 2$ (the solid black curve in Fig.\,7). This relation
is given by the equation:

\begin{equation}
\epsilon_{912} = 10^{24.9}(1+z)^{4.68} \frac{e^{-0.28z}}{e^{1.77z}+26.3}
\end{equation}

\noindent
in units of ${\rm erg\,s^{-1}Hz^{-1} Mpc^{-3}}$.

It is clear that our results fall far short of the emissivity curves produced by Giallongo et al. (2012) and Madau \& Haardt (2015), both of which suggested that high levels of 
AGN-produced emissivity might be maintained back into the reionization epoch. 
Thus, as a result of our revised photometric redshifts, and our reanalysis of the claimed X-ray detections,  our findings support the conclusion, originally reached by Haardt \& Madau (2012) (and 
more recently reaffirmed by Vito et al. 2016, Qin et al. 2017and Ricci et al. 2017), that the contribution of AGN to cosmic hydrogen ionization is essentially neglible.

Of course it is important to note that the revised model of reionization presented by Madau \& Haardt (2015) (green curve in Fig.\,7) was motivated, at least in part, by the observational 
results of Giallongo et al. (2015), and hence renewed interest in a model in  which cosmic hydrogen reionization was produced entirely by early AGN. However, Fig.\,7 shows that 
our revised estimate of AGN ionizing emissivity at $z > 5$ now lies over an order-of-magnitude below that required by such a model. This is despite the fact
that we have assumed an AGN ionizing photon escape fraction 
of unity in calculating $\epsilon_{912}$, which is almost certainly excessive (e.g. Micheva et al. 2017). Finally, we also note that models in which AGN are responsible for
HI reionization struggle to reproduce the $\tau_{HeII}$ measurements (Khaire et al. 2016; Khaire 2017).

We conclude that all available evidence continues to favour a scenario in which young galaxies reionized the Universe, with AGN playing an essentially negligible role in cosmic hydrogen 
reionization.

\section{Conclusions}

We have reinvestigated the properties of a claimed sample of 22 X-ray detected active galactic nuclei (AGN) at redshifts $z > 4$ within the GOOD-S/CANDELS field. 
While 8 of these X-ray sources were already known from existing X-ray catalogues derived from the 4\,Ms {\it Chandra} imaging within the CDF-S (Xue et al. 2011), 
the remaining 14 were identified 
by Giallongo et al. (2015). This was done using positional priors
derived from the {\it HST} $H_{160}$ images of a catalogue of 1113 sources with claimed photometric redshifts $z > 4$, and searching for clustering of X-ray photons in time coincident with
these positions.

We have re-examined the robustness of the claimed X-ray detections (within the {\it Chandra} 4\,Ms imaging), and find that, even taking an optimistic approach to 
apparent detections, only 15 of the 22 X-ray sources can be confirmed in the 4\,Ms data, with a stack of the remaining 7 sources yielding no significant detection.
We have also carefuly re-investigated the photometric redshifts of the 22 $H_{160}$ counterparts, using both galaxy and AGN SEDs, and find that many of the claimed high-redshift 
sources more likely lie at lower redshifts. Specifically, we can only confirm 12 of the 22 {\it HST} sources to lie robustly at $z > 4$.

Combining these results, we have found convincing evidence for only 7 X-ray AGN at $z > 4$ in the GOODS-S field, of which only one lies at $z > 5$. 
We have re-calculated the evolving far-UV (1500\,\AA) luminosity density produced by AGN at high redshift (using both revised number counts and a new determination
of the evolving LF) and find that it declines rapidly from $z \simeq 4$ to $z \simeq 6$, in agreement with several other recent studies of the evolving AGN luminosity function. 

Finally, we have calculated the resulting inferred high-redshift evolution of  comoving hydrogen-ionizing emissivity density ($\epsilon_{912}$), and compared 
this with other recent observations and models. We conclude that, at $ z \simeq 6$,  the contribution towards $\epsilon_{912}$ from AGN falls over an order-of-magnitude short of the 
level required to maintain cosmic hydrogen ionization, and that the redshift evolution of $\epsilon_{912}$ from AGN is better described by the model 
of Haardt \& Madau (2012) (albeit possibly boosted in overall amplitude by a factor $\simeq 2$) than by the revised model of Madau \& Haardt (2015) (which was inspired, in part, by 
the results of Giallongo et al. 2015). Our results support a scenario in which the emerging population of young galaxies reionized the Universe, with AGN 
making at most a minor contribution to cosmic hydrogen reionization.

\section*{Acknowledgments}
SP acknowledges the support of the University of Edinburgh via the Principal's Career Development Scholarship. 
This work is based in part on observations made with the NASA/ESA Hubble Space Telescope, which is operated by the Association of Universities for Research in Astronomy, Inc., under NASA contract NAS5-26555. This work is also based in part on observations made with the Spitzer Space Telescope, which is operated by the Jet Propulsion Laboratory, California Institute of Technology under NASA con- tract 1407. This work uses data taken with the Hawk-I instrument on the European Southern Observatory Very Large telescope from ESO programme: 092.A-0472

{}

\appendix

\section{X-ray images}

For each of the individual X-ray sources reported by Giallongo et al. (2015), we have revisited and inspected the soft, hard and full X-ray energy-band ($0.5-2$\,keV, $2-7$\,keV and $2-10$\,keV respectively) imaging provided by the {\it Chandra} 4\,Ms imaging of the CDF-S. To increase the chances of a significant detection we smoothed the images 
and sought a significant detection in any one of the three X-ray images. The results are shown in Fig.\,A1. Even taking an optimistic approach to detections, we can confirm only
15 of the claimed 22 X-ray sources. These 15 include the 8 previously known X-ray sources (Xue et al. 2011) plus 7 of the new sources claimed by Giallongo et al. (2015) with their
positional prior-based detection technique. This leaves 7 sources for which we can find no evidence of significant X-ray flux in any band coincident with the
position of the claimed {\it HST} counterpart 
(ID: 4285, 4952, 5501, 9713, 12130, 31334, 33073).
As discussed in the text, and shown in Fig.\,4, a stack of the X-ray imaging of these 7 sources also yields no significant X-ray detection consistent with the {\it HST} position.

\begin{figure*}
\centering
\includegraphics[width=500pt, height=630pt]{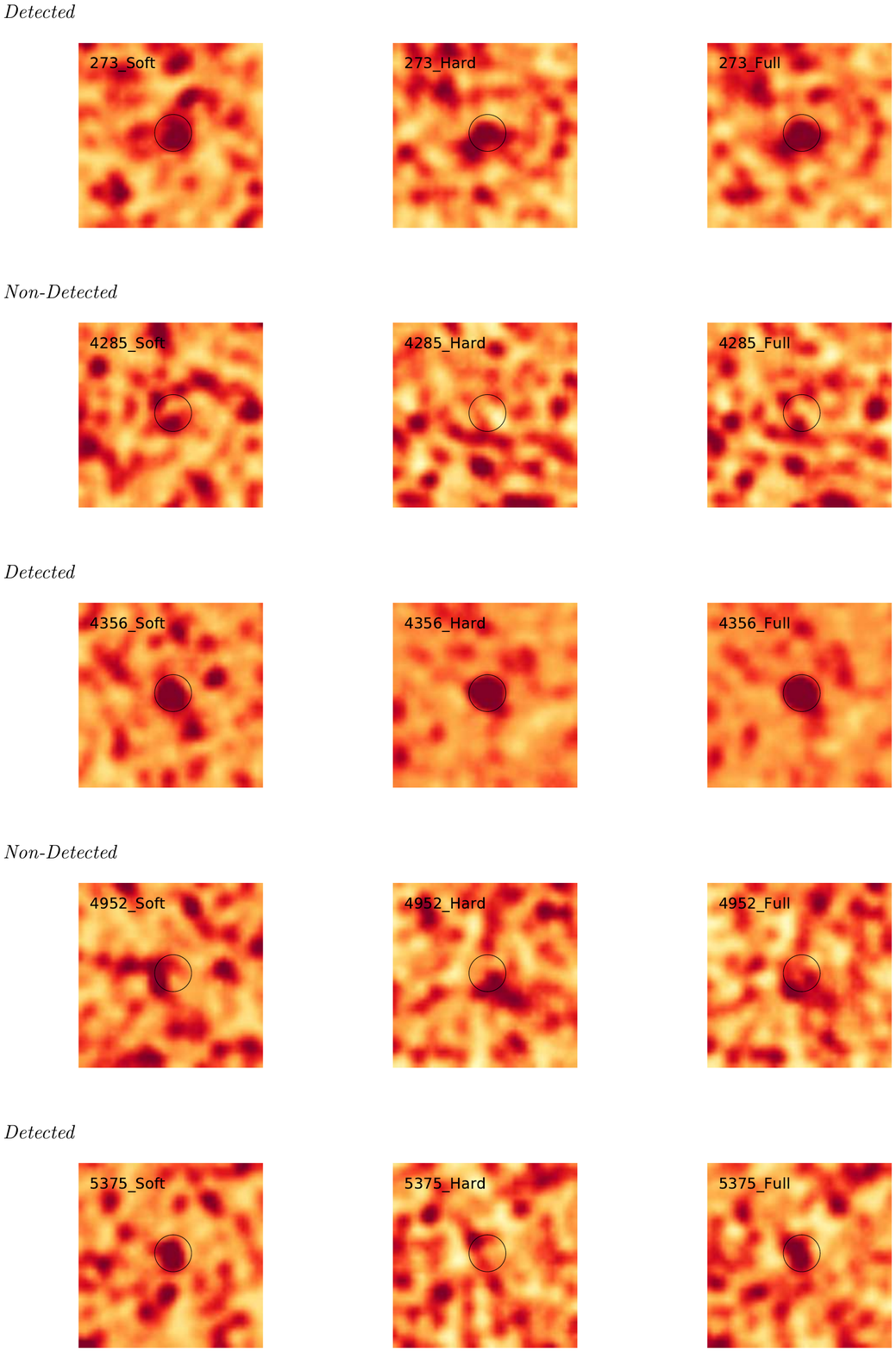}
\caption{Postage-stamp soft, hard and full-band {\it Chandra} 4\,Ms X-ray images of the 22 AGN candidates reported by Giallongo et al. (2015), as listed in Table 1.
  The imaging has been Gaussian smoothed with $\sigma = 1.5$\,arcsec, and each stamp is 20 x 20 arcsec, with North to the top and East to the left. The colour scale spans a flux-density range
  of $\pm2.5\sigma$ around the median background. 
The positions of the {\it HST} WFC3/IR optical/near-IR counterparts are marked by a circle of radius 2\,arcsec in each plot. For each `source' we indicate on the left whether there is a signficant 
detection of X-ray flux coincident with the {\it HST} counterpart, in any of the three X-ray images.}

\end{figure*}
\begin{figure*}
\centering
\includegraphics[width=500pt, height=650pt]{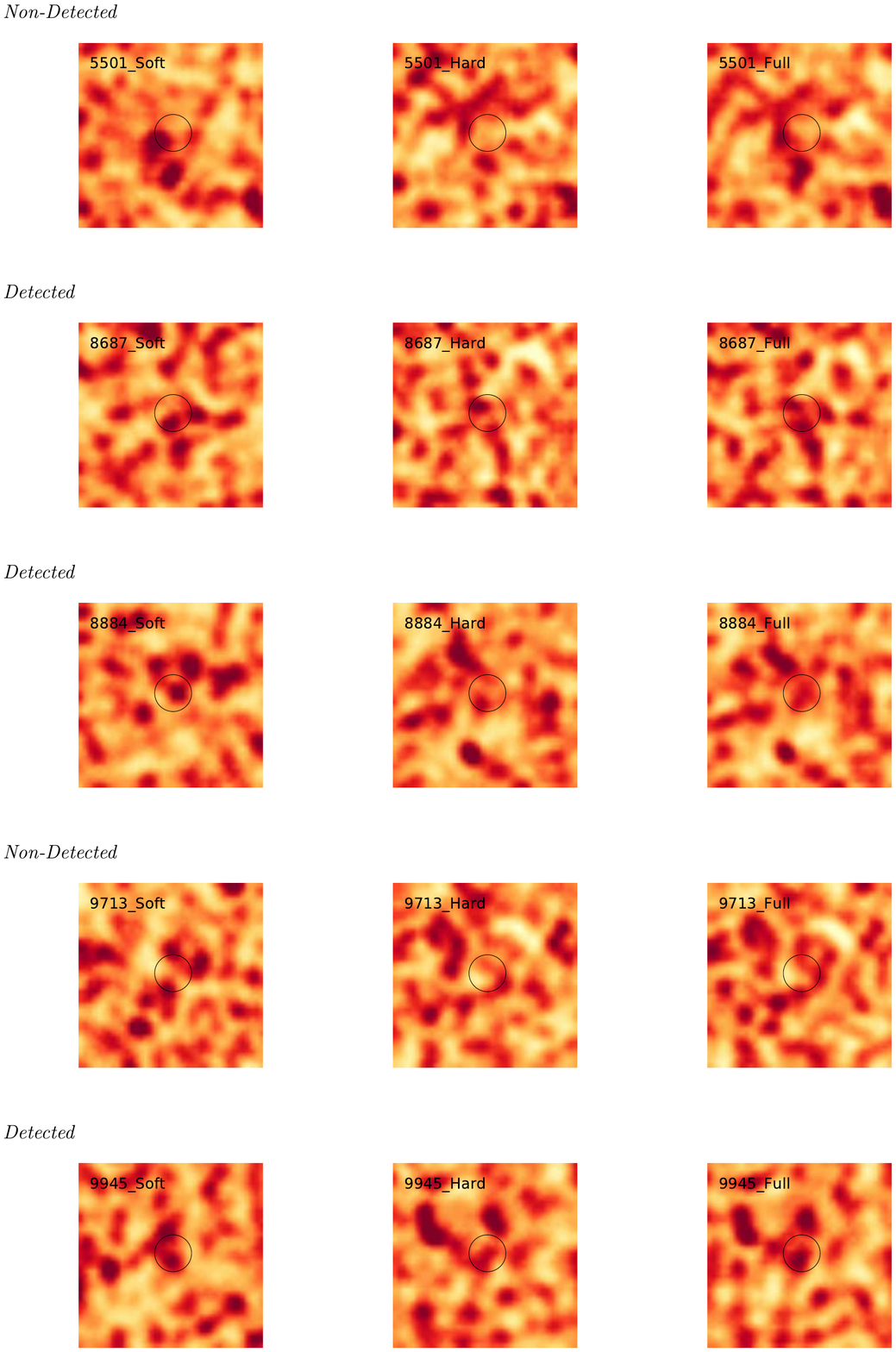}
\ContinuedFloat
\caption{(continued)}
\end{figure*}
\begin{figure*}
\centering
\includegraphics[width=500pt, height=650pt]{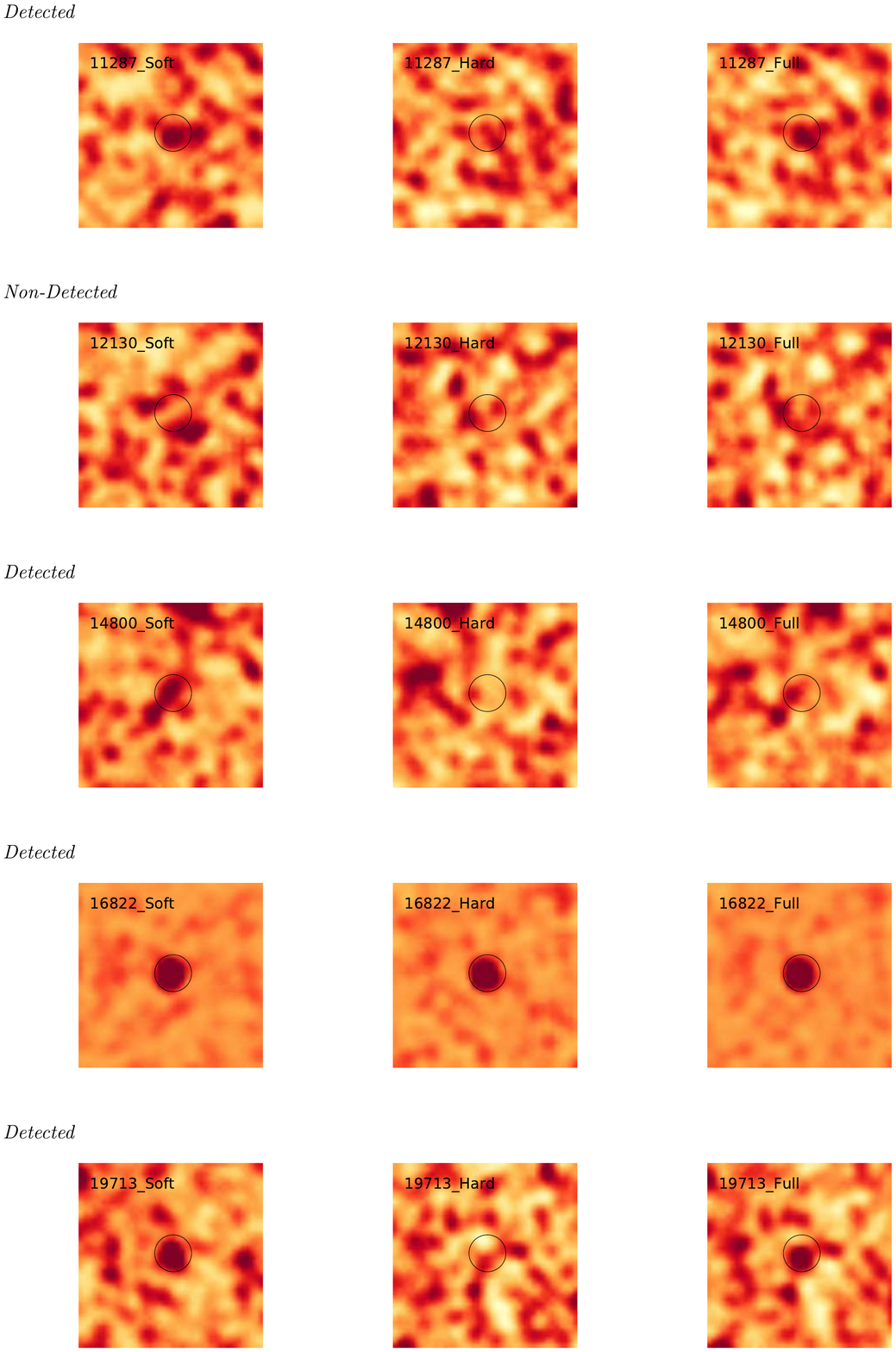}
\ContinuedFloat
\caption{(continued)}
\end{figure*}
\begin{figure*}
\centering
\includegraphics[width=500pt, height=650pt]{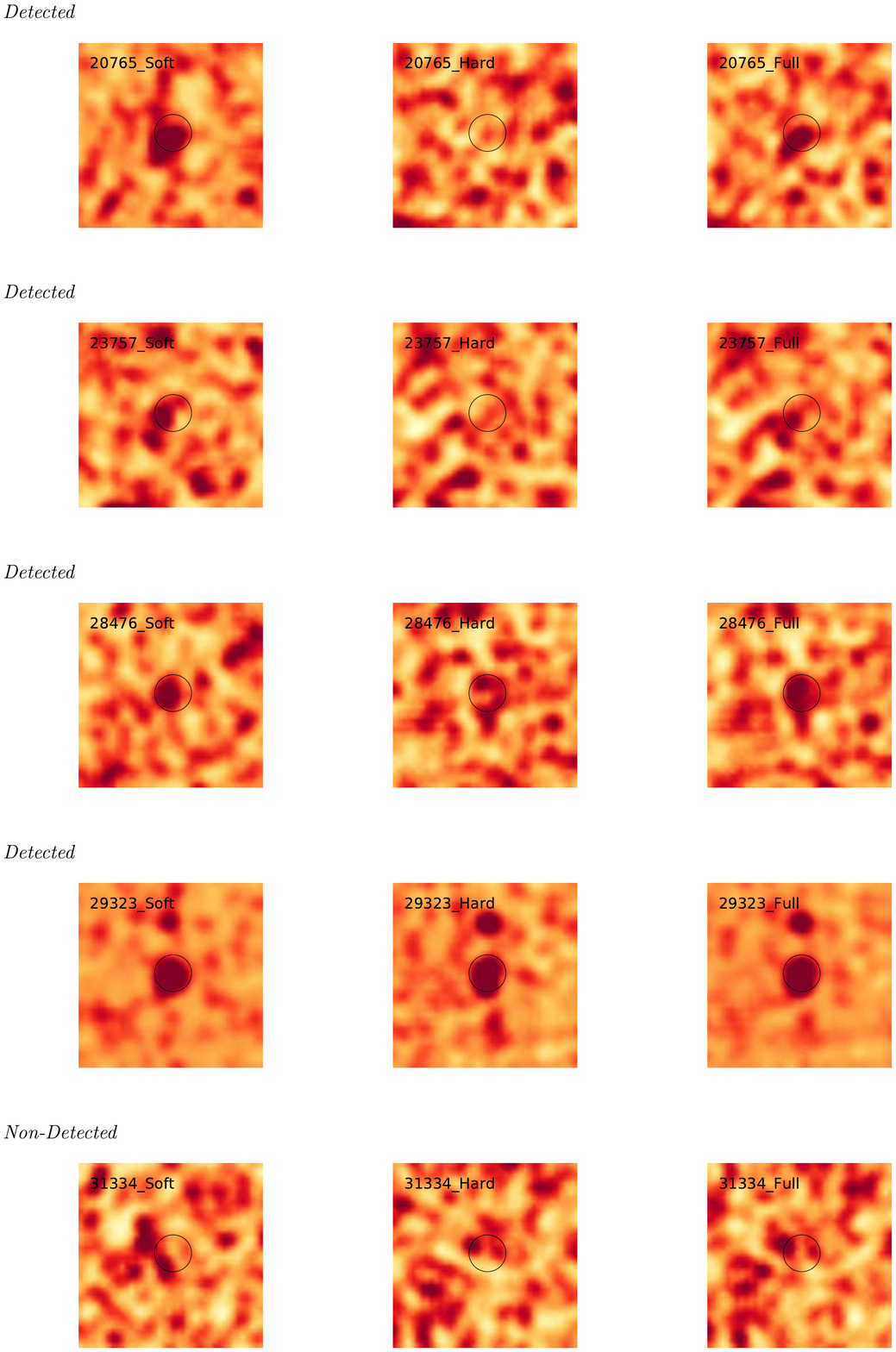}
\ContinuedFloat
\caption{(continued)}
\end{figure*}
\begin{figure*}
\centering
\includegraphics[width=455pt]{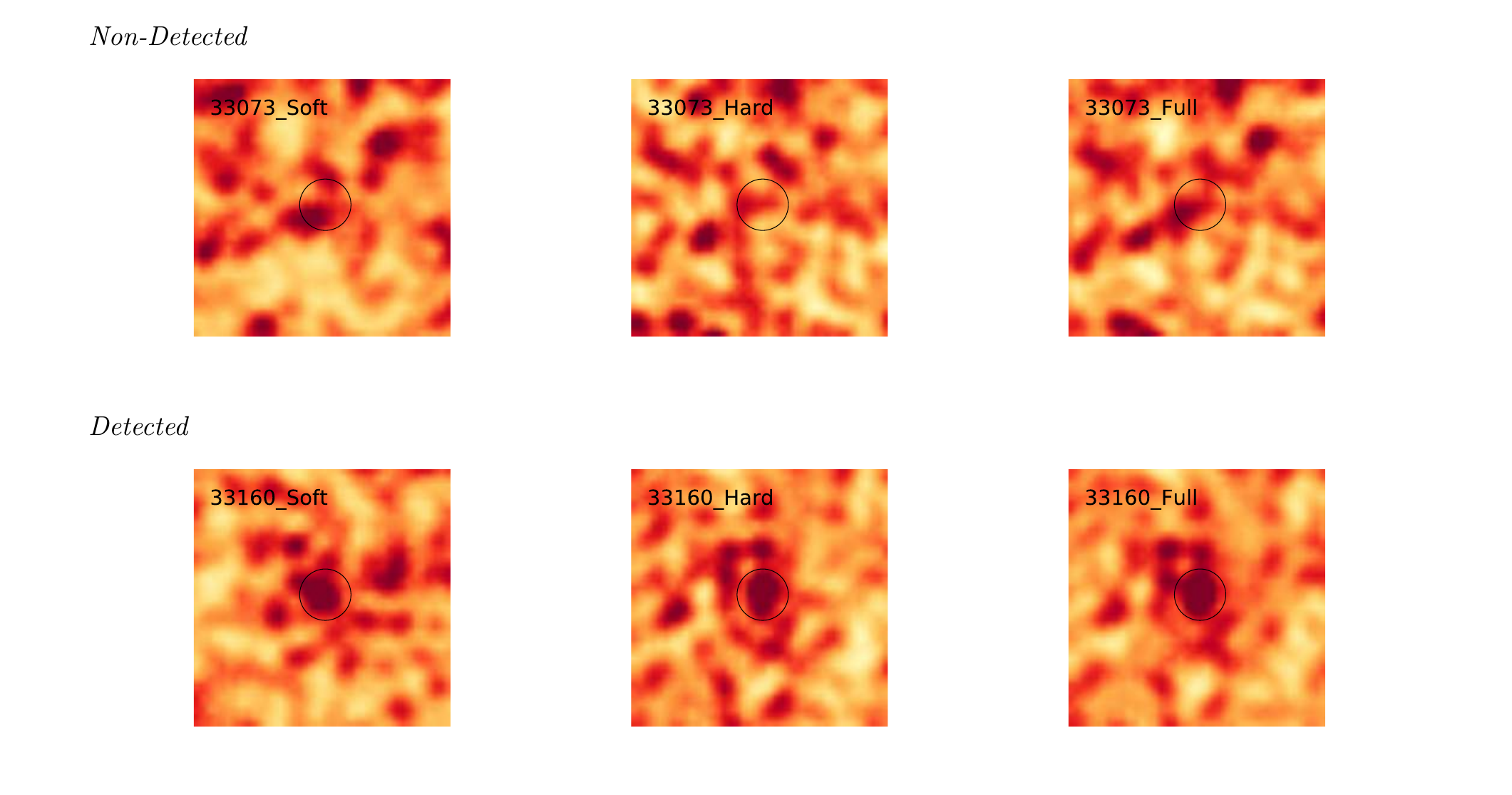}
\ContinuedFloat
\caption{(continued)}
\end{figure*}

\end{document}